\documentclass[12pt, preprint]{aastex}

\begin{document}
\title{Relativistic O {\sc viii} Emission and Ionized Outflow in NGC 4051 Measured with XMM-Newton}

\author{P. M. Ogle$^{1}$, K. O. Mason$^2$, M. J. Page$^2$, 
        N. J. Salvi$^2$, F. A. Cordova$^3$, I. M. McHardy$^4$,
        \& W. C. Priedhorsky$^5$}
\affil{$^1$ Jet Propulsion Laboratory, California Institute of Technology, MS 238-332, 4800 Oak Grove Dr., Pasadena, CA 91109}
\affil{$^2$ MSSL, University College London, Holmbury St. Mary, Dorking, 
            Surrey, RH5 6NT, UK}
\affil{$^3$ 4148 Hinderaker Hall, University of California, Riverside, CA 92521}
\affil{$^4$ University of Southampton, University Road, Southampton, UK}
\affil{$^5$ Los Alamos National Laboratory, Los Alamos, NM 87545, USA}

\email{pmo@sgra.jpl.nasa.gov}

\shorttitle{Relativistic O {\sc viii} Emission in NGC 4051}
\shortauthors{Ogle et al.}

\begin{abstract}

We present {\it XMM-Newton} RGS observations of the soft X-ray spectrum of NGC 4051, and
explore their implications for the inner accretion disk and ionized outflow in the 
active galactic nucleus. We fit the soft X-ray excess with a relativistically 
broadened O {\sc viii} recombination spectrum, including the entire line series and 
recombination continuum. This plus an underlying power law continuum provides a much better 
fit to the soft X-ray spectrum than a single temperature or disk blackbody plus power law. 
The emission line profiles, computed for a Kerr metric around a maximally rotating black hole,
reveal a sharply peaked disk emissivity law and inner radius smaller than $1.7 R_G$.
The spectrum also includes narrow absorption and emission lines from C, N, O, Ne, and 
Fe in an ionized outflow. Outflow column densities are relatively low and do not 
create significant edges in the spectrum. The small amount of absorption bolsters confidence in 
the detection of relativistic emission line features. The narrow-line emitter has a large 
(76\%) global covering fraction, leading to strong forbidden lines and filling in of the 
resonance absorption lines. We also find broad C {\sc vi} Ly$\alpha$ and very broad O {\sc vii} 
emission from the broad-line region. The narrow and broad-line regions span large ranges 
in ionization parameter, and may arise in a disk outflow. The ionized absorber has
a large ionization range which is inconsistent with pressure equilibrium in a multiphase 
medium. The mass outflow rate exceeds the accretion rate by a factor of one thousand.

~

\end{abstract}

   
\section{Introduction}

Study of relativistic emission lines in the X-ray spectra of Seyfert galaxies has the 
potential to elucidate the nature of the accretion disk and spacetime geometry near 
supermassive black holes. A very broad feature was observed by {\it ASCA} near the energy 
of the 6.4 keV Fe K$\alpha$ line in MCG-6-30-15 \citep{t95} and other Seyfert galaxies
\citep{ngm97b}. Convincing examples measured by {\it XMM-Newton} include MCG-6-30-15 
\citep{wrb01} and Mrk 766 \citep{pmc01,mbo03}. The current best explanation for this feature 
is Doppler and gravitationally broadened Fe K-shell emission from an accretion disk which 
extends close to the innermost circular orbit of a supermassive black hole \citep{frs89}.

The recent discovery of relativistic emission lines in the {\it soft} X-ray spectra of MCG-6-30-15 
and Mrk 766 \citep{br01} may help to reveal the ionization structure and elemental abundances 
in the accretion disk. The O {\sc viii}, N {\sc vii}, and C {\sc vi} Ly$\alpha$ emission lines 
probe high ionization states which may arise in the photoionized surface of the disk. 
This is complementary to studies which have indicated emission from highly ionized Fe 
in these same galaxies \citep{wrb01,mbo03}.
 
Ionized accretion disk models have been compared to the observations,
but have yet to be explicitly fit to grating spectra. The predicted emission line equivalent
widths depend primarily on the vertical ionization structure and abundances in the top layer of the 
accretion disk. Both constant density \citep{brf02} and hydrostatic balance \citep{nkk00} have 
been used to set the density profile of the disk. Together with the disk illumination law,
this determines the ionization structure. Constant density slab models predict O {\sc viii}
Ly$\alpha$ equivalent widths of 30-70 eV for a disk with solar abundances \citep{brf02}. The 
observed O {\sc viii} equivalent widths in MCG-6-30-15 and Mrk 766 are 2-5 times these predicted 
values. However, the optical depth of the photoionized skin can be increased by raising the ratio of
disk flux to X-ray flux, explaining the observed O {\sc viii} Ly$\alpha$ equivalent widths 
\citep{skb03}.

More problematic are the very strong N {\sc vii} Ly$\alpha$ emission lines and lack of Fe L-shell 
lines observed in MCG-6-30-15 and Mrk 766. The observed equivalent widths require N/O abundance ratios 
which are several times solar. Radiative transfer effects may be important for explaining the anomalous
emission line ratios. It is possible to pump the N {\sc vii} Ly$\alpha$ emission line with O {\sc viii}
Ly$\alpha$ photons via Bowen resonance fluorescence \cite{s03}. The lack of relativistic Fe L-shell 
lines may be explained by a large O {\sc viii} edge opacity, while the O {\sc viii} Ly$\beta$ line 
could be suppressed by resonance scattering \citep{skb03}. It is also essential to model the effects of
an overlying ionized absorber component \citep{mbo03}.

An alternative model has been proposed to explain the soft X-ray spectral features in 
MCG-6-30-15 and Mrk 766 by absorption edges from neutral Fe in dust instead of relativistic 
emission lines \citep{loc01}. The apparently reddened optical-UV continuum, strong X-ray absorption,
and weak UV/X-ray flux ratio in  MCG-6-30-15 \citep{rwf97} supports the notion of a dusty ionized
absorber. In this model, the dust must be embedded in a highly ionized medium in order to avoid 
strong edges from neutral He, C, and N which would obliterate the X-ray spectrum at the column 
densities inferred from the optical reddening. However, \cite{mbo03} and \cite{skb03} argue 
against the dusty ionized absorber model based on the lack of strong Fe {\sc i} resonance 
absorption features which should attend the putative Fe L-shell edges. The lack of intrinsic 
O {\sc i} K-edges in the {\it XMM-Newton} RGS spectra of MCG-6-30-15 and Mrk 766 is further evidence
against silicate dust.

Low resolution observations of NGC 4051 with EXOSAT \citep{mgd95} and ASCA \citep{g96}
indicated the presence of variable absorption features at 0.6-0.9 keV. These were interpreted
as O {\sc vii} and O{\sc viii} edges from an ionized absorber along the line of
sight to the X-ray continuum source. 
Grating observations of NGC 4051 and several other sources reveal a more complicated picture. 
Resonance absorption lines and unresolved transition arrays (UTAs) are responsible for much
of the soft X-ray spectral complexity in Seyfert galaxies. In hindsight, it is obvious that
resonance absorption lines can be much stronger and appear at much lower column densities
than photoionization edges. High resolution {\it Chandra} and {\it HST} grating spectra 
of the ionized X-ray and UV absorbers in NGC 4051 are reported by \cite{cbk01}. There are 
multiple blue-shifted absorber components, and re-emission is seen from narrow X-ray 
lines.

NGC 4051 is classified as a narrow-line Seyfert 1 galaxy (NLS1). In order to avoid
confusion when talking about the different emission line regions, we refer to
them as follows. The narrow-line region (NLR) has the lowest velocity
and turbulent velocity (FWHM$<5 \times 10^2$ km s$^{-1}$). The broad-line region (BLR) 
has larger velocity width ($5 \times 10^2-3\times 10^4$ km s$^{-1}$). The 
relativistic-line region (RLR) has the largest velocity width ($>3\times 10^4$ km 
s$^{-1}$) and is also affected by gravitational redshift. Note that this 
classification is based purely on velocity, which increases closer to the central 
black hole. We avoid a classification based on density, for as we shall show, both the
NLR and BLR (and perhaps RLR) encompass a large range of ionization parameter and 
density.

In this paper, we present the {\it XMM-Newton} RGS spectrum of NGC 4051 (z = 0.00234). 
Our long (122 ks) observation reveals unprecedented detail in the soft X-ray spectrum. 
We interpret a very broad spectral feature as relativistic O {\sc viii} emission from 
the inner accretion flow around the central black hole. We also fit models of the ionized 
absorber and emitter spectrum, including inner-shell absorption from a number of ions.
    
\section{Observations}
    
We observed NGC 4051 for a full orbit with {\it XMM-Newton} on 16-17 May
2001. The OM and EPIC light curves are reported by \cite{mmp02}
and the EPIC spectral variability is analyzed by \cite{salv03} and \cite{spw03}. The
X-ray flux varied significantly during the observation with a period of
particularly low flux approximately half way through \citep{mmp02}. During the last 
14 ks of the observation the particle background increased dramatically. We 
have excluded these data from the analysis. The spectrum during 
the 8.4 ks period of lowest flux ("low state", 0.3-2.0 keV EPIC PN+MOS count rate 
$< 8$ count s$^{-1}$) was reduced separately from the remaining 99 ks ("high state") 
of the observation. 

The RGS spectrum (Fig. 1) was reduced using the {\it XMM-Newton} SAS v5.2. 
The response matrices were corrected for imperfections in the effective area calibration 
using the data/model ratios from a power law fit to the RGS spectra of Mrk
421. After this correction the individual first and second order spectra
from RGS1 and RGS2 were added together and the response matrices were
combined accordingly. The absolute effective area of RGS is calibrated to 
better than 5\% \citep{dh01}.

The spectral resolution of RGS in first order is FWHM$= 72$ m\AA, and 
the wavelength calibration is accurate to 8  m\AA\ \citep{dh01}. These 
correspond to FWHM$= 620$ km s$^{-1}$ and $\Delta v=69$ km s$^{-1}$ at 35 \AA. 
The spectral bins range from 20-40 m\AA\ in width from the low to high wavelength
ends of the spectra, so we oversample by a factor of 2-4. The high-state
spectrum has 50-570 photons/bin, while the low-state spectrum has 2-22 
photons/bin. There are enough photons/bin in the high-state spectrum to use 
$\sqrt N$ Poisson errorbars and fit our models using the $\chi^2$ statistic. For the low 
state, we use the C-statistic \citep{c79} and plot asymmetric Poisson errorbars 
\citep{g86}.

The combined {\it XMM-Newton} EPIC, RGS  and Optical Monitor (OM) measurements \citep{salv03}
constrain the photoionizing portion of the NGC 4051 spectral energy distribution 
(SED, Fig. 2). The UV flux points at 2100, 2300, and 2900 \AA\ were computed from the mean 
count rates in the UVW2, UVM2, and UVW1 bandpasses and corrected for Galactic reddening 
\citep{salv03,mmp02}. Additional radio, IR, and optical flux points were 
obtained from the NASA Extragalactic Database (NED). The soft X-ray slope of $\Gamma=2.4$
is from our best fit model to the high-state RGS data. The hard X-ray slope of $\Gamma=1.5$ is 
from the best fit PL to the high-state EPIC PN data. We assume that the SED has 
a high-energy cutoff above 100 keV. Integrating over the SED, the bolometric luminosity is 
$L_\mathrm{bol}=2.7 \times 10^{43}$ erg s$^{-1}$ and the ionizing (13.6 eV-13.6 keV)
luminosity is $L_\mathrm{ion}=4.1 \times 10^{42}$ erg s$^{-1}$ during the high-state.
   
We retrieved archival {\it Chandra} HETGS observations taken during 2000 April 24-25 for comparison 
with the 2001 May {\it XMM-Newton} data set. The Medium Energy Grating (MEG) data are most useful 
for assessing the nature of the soft X-ray excess, and have been analyzed in detail by 
\cite{cbk01}. We reprocessed the MEG $\pm 1^{\mathrm st}$ order spectra from the Level 1 event file 
using CIAO 3.0.1 and the associated CALDB 2.23 calibration database. We combined events from the 
entire 82 ks of the observation to get the mean spectrum. The two first order spectra were 
added and the resultant spectrum was rebinned to 0.05 \AA\ to improve the statistics at
long-wavelengths. This yielded 10-270 counts per bin, enabling us to use Poisson errorbars and
to use $\chi^2$ as a fit statistic. The unbinned MEG spectra are too noisy to allow
a meaningful analysis at the full 0.023 \AA\ resolution of the spectrograph.
  
\section{Spectral Models}

\subsection{Continuum}
   
The high state RGS spectrum of NGC 4051 is far from a simple power law
(Fig. 1). There is a large soft X-ray excess over the underlying soft power law, apparent as a 
broad bump from 13-32 \AA. The spectrum is punctuated by many narrow absorption lines, 
but there are no deep absorption edges. There are several narrow emission lines which 
contribute a small percentage of the soft excess seen in the lower resolution 
EPIC spectrum \citep{spw03}. The nature of the curvature in the soft spectrum is not obvious, 
and has previously been ascribed to blackbody emission \citep{g96,cbk01}.  

We use the IMP\footnote{IMP and its documentation are available at 
\url{http://xmmom.physics.ucsb.edu/\~\ pmo/imp.html}} (v1.1) spectral fitting code
developed at UCSB \citep{bo03, ob03} to fit the high-state RGS spectrum. All 
fits include Galactic absorption with column density 
$N_\mathrm{H} = 1.32 \times 10^{20}$ cm$^{-2}$ \citep{dl90}. They 
also include narrow absorption and emission from the intrinsic ionized outflow, which 
we discuss in detail below. We adopt a distance of 9.35 Mpc, derived from the host 
galaxy kinematic redshift assuming $H_0=75$ km s$^{-1}$ Mpc$^{-1}$, to convert from 
flux to luminosity.  Unless otherwise stated, single parameter error estimates are for 90\%
confidence.

We first try fitting a power law plus single temperature blackbody to the RGS spectrum (Fig. 1a). 
This model gives a formally poor fit ($\chi^2/DF = 1.56$). The blackbody peak is broader 
than the observed soft excess. This is compensated by a deep O {\sc vii} edge at 17 \AA\ which 
appears to be overestimated by the model. The best fit temperature 
for the blackbody is $kT = 0.14$ keV, and the underlying power law has 
photon index $\Gamma = 2.2$. These are similar to the results of \cite{cbk01}, 
who suggest that residuals to their blackbody fit represent additional spectral complexity
from absorption.

A power law plus multicolor disk blackbody gives a worse fit ($\chi^2/DF = 1.66$).
The best-fit black hole mass for this model is $5\times 10^3 M_\odot$. The inner
radius of the accretion disk is $1.3 R_\mathrm{G}$, and the accretion rate is 60\% of the Eddington
rate. While the model roughly fits the shape of the soft X-ray continuum, it underpredicts the 
UV flux and bolometric luminosity by a large amount (Fig. 2). It is inconsistent with 
reverberation studies which yield a much larger black hole mass of $\sim10^6 M_\odot$ 
\citep{p00,sun03}. For such a large central mass, a disk blackbody should peak in the far-UV, 
not the X-ray band. 

It is unlikely that the soft excess is the Wien tail of a Comptonized blackbody. We
tested this possibility with the IMP {\it comptt} model \citep{tit94}. The observed spectral 
slope isn't nearly steep enough, especially at the long wavelength end of the RGS spectrum.
The shape of such a model is so far off, that we do not attempt to fit it to the
data. A Wien tail absorbed by neutral carbon in a dusty ionized absorber appears to be ruled
out since there is no other indication of dust absorption in the spectrum of NGC 4051.

For the sake of completeness, we also try a power law plus optically thin bremsstrahlung 
model (Fig. 1b). This gives a worse fit than the blackbody model ($\chi^2/DF = 1.69$). The 
best fit plasma temperature is $kT=0.83$ keV, and the emission measure is $9.8\times 10^{64}$
cm$^{-3}$. Because the roll-over in the bremsstrahlung spectrum is not steep enough
to match the observed spectral bump, the model compensates by overestimating the O {\sc vii} 
edge depth.
   
Conventional continuum emission processes can not explain the observed spectral bump,
so we turn to discrete atomic processes. Unlike MCG-6-30-15 \citep{loc01}, the 
spectral break at 17 \AA\ is much too broad to be explained by O {\sc vii} K and 
Fe {\sc i} L photoionization edges from the ionized absorber. As we show below, the observed 
column densities of O {\sc vii} and Fe {\sc i} in the ionized absorber are too small to 
give appreciable edges.  

\subsection{Relativistic Lines}
 
Next we model the soft X-ray spectrum with a power law plus
relativistic O {\sc viii} Ly$\alpha$ emission line. We assume line profiles 
from an accretion disk which may extend to the inner stable circular orbit of a 
maximally rotating Kerr black hole \citep{l91}. The disk emissivity as a function of
radius is assumed to follow a power law $R^{-q}$. The inner radius is allowed to
vary, and the outer radius is fixed at $R_\mathrm{o} = 400 R_\mathrm{G}$, 
where $R_\mathrm{G}=GM/c^2$. This model (REL1, $\chi^2/DF = 1.48$) does better than the 
blackbody model, but  underestimates the flux in the 13-17 \AA\ band (Fig. 3d, e). 
The extent of the red wing of the line is parameterized by an inner radius
of $R_\mathrm{i}<1.5 R_\mathrm{G}$, and a radial emissivity law with power law index of 
$q_1 =5.39$. The disk  inclination is $i_1=52.7\arcdeg$.

Our best model (REL2) of the spectrum (Fig. 3a, b, c) includes 
relativistic emission from the entire O {\sc viii} emission line series and radiative 
recombination continuum (RRC). The line and RRC strengths are computed using the {\it ionemit} 
model, which is described in \S 3.3. We assume that the O {\sc viii} ion is photoionized, with 
an electron temperature of $T_\mathrm{e}=2\times 10^6$ K (consistent with $\log \xi=3.4$, see \S 4.2). 
For simplicity, we assume that all emitted line photons escape directly and do not treat their
radiative transfer. Radiative transfer effects could alter the observed line ratios. For example,
Ly$\beta$ photons may be down-converted to Ly$\alpha$ if the emission region is optically thick.

The REL2 model has $\chi^2/DF = 1193/901 = 1.32$, a significant improvement over REL1 ($\Delta 
\chi^2=201$ for one additional degree of freedom). The combined Ly$\beta$ plus higher 
order lines account for the gradual steepening of the spectrum in the 13-17 \AA\ band. 
We map cross-sections of the $\chi^2$ surface in ($R_\mathrm{i}$, $q$, $i$, $\Gamma$) parameter space 
(Fig. 4), parameters which have the largest influence on the relativistic line shape. There is a 
well-localized minimum in each of these plots, demonstrating that we can accurately constrain 
disk parameters. We adopt the $3\sigma$ (99.7\%)  confidence region for variation in two parameters 
as a conservative bound on these parameters, given possible systematic uncertainties in the 
underlying continuum shape and ionized absorber-emitter models.

The breadth of the relativistic O {\sc viii} line profiles is parameterized by an inner disk
radius of $R_\mathrm{i}<1.7 R_\mathrm{G}$ at 99.7\% confidence (Fig 4a, b, c). As far as we are aware, 
this is the closest distance to the event horizon of a supermassive black hole ever measured. 
If the line emission comes from outside the innermost stable circular orbit (ISCO), it implies a 
black hole spin parameter of $j>0.973$, approaching the maximal value. Emission from {\it inside} 
the ISCO is not excluded, but would require an extension of our model. The greatest uncertainty
on the disk inner radius comes from modeling the underlying continuum. Decreasing the power law
index by 0.1 dex from the best fit value increases $R_\mathrm{i}$ from 1.24 to 1.6 $R_\mathrm{G}$, 
and still yields and acceptable fit (Fig. 4c).
 
There is some degeneracy between the disk emissivity index and the disk inclination (Fig. 4c),
since both affect the energy of the blue shoulder of the relativistic line profile. However, the
emissivity index is still well constrained to be $q_2 =5.0^{+0.5}_{-0.2}$, at 99.7\% 
confidence for variation in 2 parameters. This means that nearly all ($>99\%$) of the 
O {\sc viii} emission comes from within 6 $R_\mathrm{G}$ and the half-light radius is 
$1.6 R_\mathrm{G}$ (for $R_i = 1.24 R_\mathrm{G}$). A steep radial emissivity law ($q=3.5$) is 
predicted for a black hole accretion disk with non-zero torque at the ISCO \citep{ak00}. The 
relativistic lines in Mrk 766 and MCG-6-30-15 \citep{br01} have emissivity indices of 
$q = 3.6-3.8$ and half-light radii of $\sim 1.9 R_\mathrm{G}$, which are similar to this 
prediction. The radial emissivity index for the relativistic O {\sc viii} line in NGC 4051 
is much steeper, implying that it originates from a narrower range of radii.

The energy of the blue shoulder of the relativistic O {\sc viii} Ly$\alpha$ line profile is 
sensitive to and increases with disk inclination. The best-fit disk inclination for NGC 4051 is 
$i_2=48^{+4}_{-3} \arcdeg$, again at 99.7\% confidence for variation in 2 parameters. This
value is correlated with $q_2$, with steeper values of the emissivity index yielding larger disk
inclination. We have not included smearing from Compton scattering, which could lead to an 
overestimate of the energy of the blue shoulder and hence the inclination. It is also possible 
that a flare or warp to the inner disk or other time dependent structure will affect the 
inclination. With these caveats in mind, we find that the disk inclination in NGC 4051 is 
significantly different from the inclinations ($36\arcdeg$, $40\arcdeg$) inferred from the 
relativisitic line profiles in Mrk 766 and MCG-6-30-15 \citep{br01}. This is to be expected from
AGN viewed at different angles. The disk inclination is also consistent with the 
$48 \pm 5\arcdeg$ inclination of the conical NLR outflow in NGC 4051, derived from [O {\sc iii}]
imaging and spectroscopy \citep{chs97}. This alignment is remarkable given the much larger size 
scale of the NLR and may indicate an intimate connection between the NLR and accretion disk.

The relativistic O {\sc viii} Ly$\alpha$ flux is $F = 3.5\pm 0.1 \times 10^{-3}$ ph 
s$^{-1}$ cm$^{-2}$ and the equivalent width is $88 \pm 3$ eV. Using a 
curve-of-growth analysis, the observed EW requires an O {\sc viii} column density of 
$\sim10^{19}$ cm$^{-2}$ for a reasonable turbulent width of $b\leq 10^4$ km s$^{-1}$ 
\citep{sdb03} and a covering fraction of unity. The O {\sc viii} Ly$\alpha$ line is dominated
by its damping wings in this region of parameter space. The sum of relativistic O {\sc viii} 
line series equivalent widths is $164 \pm 5$ eV. The model Ly$\alpha$ to Ly$\beta$ flux ratio 
is 3.8-4.7, and the ratio of RRC to Ly$\alpha$ is 0.19-0.23. The exact ratios depend on the 
O {\sc viii} column density and Doppler parameter in the accretion disk. These are not 
well constrained by the model fit, owing to severe blending of higher order lines and RRC.

There are no relativistic lines from N {\sc vii} or C {\sc vi} in the NGC 4051
spectrum. We put upper limits (at 90\%  confidence) of $4 \times 10^{-4}$ and 
$6 \times 10^{-4}$ ph s$^{-1}$ cm$^{-2}$ on the Ly$\alpha$ fluxes of these two ions, 
assuming all of the relativistic line profiles are the same. The photon flux ratios with 
respect to O {\sc viii} Ly$\alpha$ are then $< 0.09$ for N {\sc vii} Ly$\alpha$ and 
$< 0.14$ for C {\sc vi} Ly$\alpha$. As we discuss in \S 4.2, these limits are consistent
with solar abundances in a highly ionized plasma. In comparison, the ratios observed in Mrk 
766 (1.0 and 0.8 , respectively) are more difficult to explain \citep{mbo03,s03}.

The underlying soft power law continuum in the REL2 model has $\Gamma = 2.35$ 
(for $R_\mathrm{i}=1.24 R_\mathrm{G}$)
and a normalization of $3.85 \pm 0.02 \times 10^{-4}$ ph s$^{-1}$ cm$^{-2}$ \AA $^{-1}$ 
(at 1 \AA). This yields a high-state 0.3-10 keV luminosity of $L_\mathrm{x}=5.9 \times 10^{41}$
erg s$^{-1}$ from the soft power law. We note that other shapes for the soft X-ray continuum are
conceivable, and the chosen continuum may have an affect on the relativistic line parameters. 
However, lacking a physical model for the continuum, a power law is often used and is consistent
with the expectation of a Compton-scattered accretion disk spectrum. 

We try adding an intrinsic neutral absorber to the REL2 model and find a best fit
value of $N_\mathrm{H,int} = 1.1\pm 0.4 \times 10^{19}$ cm$^{-2}$ and no significant 
improvement in the fit. This is consistent with the result of \cite{cbk01}, who find an 
upper limit of $N_\mathrm{H,int} \sim 10^{20}$ cm$^{-2}$ from their {\it Chandra} HETGS 
observation.
  
\subsection{Ionized Outflow}
    
The RGS spectrum has strong, narrow K-shell absorption lines from C {\sc v}-{\sc vi}, 
N {\sc vi}-{\sc vii}, O {\sc vii}-{\sc viii}, and Ne {\sc ix} (Fig. 5). We detect weaker
L-shell absorption lines from Fe {\sc xvii}-{\sc xx} and inner-shell absorption 
from N {\sc iv}-{\sc v}, and O {\sc iv}-{\sc vi}. Narrow H-like, He-like, and Fe L-shell 
lines are also seen in emission. O {\sc vii} and Ne {\sc ix} forbidden emission lines are 
particularly strong, since they are not absorbed. The large equivalent width of O {\sc vii} f 
indicates a large global covering fraction for the X-ray NLR, estimated below. We infer that 
re-emission is important and fills many of the absorption lines, making them appear weaker.

The IMP models {\it ion} and {\it ionemit} are used to simultaneously model the ionized
absorber and emitter spectra, ion by ion. Absorption lines are modeled by Voigt
profiles, with cross-sections from \cite{v96b} and \cite{bn02}. High-order lines
are extrapolated using the hydrogenic approximation. Absorption edges are modeled using 
\cite{v96a} cross sections. The emission line model {\it ionemit} is described in detail 
by \cite{obc03}. Atomic data are calculated with the Fast Atomic Code \citep[FAC,][]{g02}. 
We assume optically thin clouds illuminated by a central point source. Level populations 
are calculated from equilibrium rate equations including terms for photoionization, 
recombination (RR and DR), photoexcitation, and cascades.

We fit for the line-of-sight covering fractions of both the continuum source
and NLR, which we assume are the same for all ions in the absorber. 
The column densities of the ions in the emitter and absorber are allowed to vary
independently. We assume electron temperatures for the emitter in the range
$3-9\times 10^4$ K, derived using XSTAR\footnote{XSTAR documentation is available at
\url{http://heasarc.gsfc.nasa.gov/docs/software/lheasoft}} 
\citep{km82,k02} and the observed SED (Fig. 2). The emission is normalized by 
$f_\mathrm{c}L_\mathrm{x}$, the product of the covering 
fraction and 0.3-10 keV luminosity of the soft X-ray power law. We use the best-fit 
O {\sc vii} f redshift of $z = 0.0019 \pm 0.0003$ and Doppler parameter of 
$b = 220 \pm 80$ km s$^{-1}$ for the emitter. The redshift and Doppler parameter of 
the absorber are fit using all observed lines. We measure velocity with respect to the 
host galaxy H {\sc i} kinematic redshift of z=0.00234 \citep{vs01}.
     
The mean velocity of the O {\sc vii} emission region is $v = -130 \pm 90$
km s$^{-1}$, consistent with the results of \cite{cbk01}. Allowing for the 
uncertainty in the RGS wavelength scale, this is consistent with the rest frame of the 
host galaxy. The best-fit  absorber redshift is $z = 0.00100^{+0.00003}_{-0.00007}$, 
corresponding to a velocity of $v = -402^{+9}_{-20}$ km s$^{-1}$ in the rest frame of the 
host galaxy. The best-fit absorber Doppler parameter is $b = 212^{+9}_{-6}$ km s$^{-1}$, 
which is not directly resolved by RGS. We are able to measure very accurate mean 
values for the absorber redshift and Doppler parameter by including all of the observed 
absorption lines in our model fit. However, parameters for the absorber may depend on 
ionization state. They are also sensitive to redshifts and velocity widths of the resonance 
emission lines. The mean absorber parameters are consistent with the system observed at 
$v = -600 \pm 130$ km s$^{-1}$  s$^{-1}$ ($b = 460 \pm 280$ km s$^{-1}$) with {\it Chandra} 
HETGS \citep{cbk01}. The outflow speed we measure may be slightly lower because our model takes 
into account filling in of absorption lines by emission lines.  

We do not detect the high velocity system at $v = -2340$ km s$^{-1}$ detected in the co-added 
HETG spectra of H-like ions \citep{cbk01}. It is likely that such a summed absorption 
spectrum is contaminated by lines from other ions, especially Fe L-shell lines. This 
procedure is also suspect because Voigt profiles should be added logarithmically, not linearly. 
RGS does not have as much sensitivity as Chandra to the high ionization lines at short 
wavelengths, so we can not make a definitive assessment of the presence of the high velocity 
system in the RGS spectrum. A higher S/N Chandra observation is needed to study individual line 
profiles.

The best-fit absorber column densities are given in Table 1. They range from 
$2 \times 10^{15}$ cm$^{-2}$ for N {\sc iv} to $4 \times 10^{17}$ cm$^{-2}$ for 
O {\sc viii}. We calculate peak ionic abundances using the XSTAR photoionization
code and observed SED (Fig. 2), and use these to derive equivalent hydrogen column densities of 
$3 \times 10^{19}$  to $3 \times 10^{21}$ cm$^{-2}$. There is a trend of increasing 
column density with increasing ionization parameter (Fig. 6). The N {\sc ii} ion 
(below the ionization range plotted in Fig. 6) has an anomalously large column density 
(Table 1). We expect a smaller column density than N {\sc iii}, which is undetected, so we are 
suspicious of the N {\sc ii} 1s-2p line identification.

The O {\sc vii} and O {\sc viii} column densities in the absorber correspond to K-edge
optical depths of $\tau=7.3\times10^{-2}$ and $3.8\times10^{-2}$, respectively.
As we noted above, these weak edges have little effect on the shape of the continuum 
spectrum. In comparison, \cite{cbk01} find apparent oxygen edge depths of 3-4 times 
greater in their analysis of the HETG spectra. However, they conclude that these are 
inconsistent with an absorption line curve-of-growth analysis which indicates smaller column 
densities. We surmise that their overestimate of edge depths is due to an inaccurate 
continuum model and neglect of Fe L-shell absorption. The upper limit on the Fe {\sc i} column 
density in the RGS spectrum (Table 1) corresponds to $\tau<3.0\times 10^{-3}$ at the Fe I L 
edge, and there is no other indication of silicate or iron oxide dust 
absorption.

The best fit emitter column densities are given for H-like and He-like ions in Table 
1. They lie in the range  $4 \times 10^{15}$  to $3 \times 10^{17}$ cm$^{-2}$.
Except for C {\sc v} which is not detected, they are within a factor of 2.6 of the
corresponding absorber column density. This is reassuring, and is consistent with
the emitter and absorber coming from the same system of NLR clouds. We do not
expect them to be identical, since there are likely to be fluctuations in NLR column
density. Absorption probes a pencil beam through the NLR while emission is 
integrated over the entire NLR.

We fit simultaneously for covering fraction as well as column density and 
Doppler parameter. The best-fit line-of-sight covering fraction of the NLR emitter by the 
absorber is pegged at $f_\mathrm{los}=1.0^{+0}_{-0.02}$, indicating that the bulk of the NLR lies 
interior to the absorber. This, plus the small difference in redshift between the absorber
and emitter accounts for the partial filling in of the resonance absorption lines
and the apparent absence of strong 2p-1s (r) resonance emission lines from N {\sc vii}, 
O {\sc vii}, O{\sc viii}, and Ne {\sc ix}. The best-fit line-of-sight covering 
fraction of the continuum source by the absorber is also $>0.98$. This rules out any  
significant source of extended soft X-ray continuum. 

The 0.3-10 keV ionizing luminosity intercepted by the NLR emitter is 
$f_\mathrm{c} L_\mathrm{x}=4.5\pm 0.2 \times 10^{41}$ erg s$^{-1}$. Dividing by the soft power law 
continuum luminosity gives an estimate of $f_\mathrm{c}=0.76 \pm 0.03$ for the NLR global covering 
fraction. This accounts for the large equivalent widths of the narrow forbidden emission lines. We 
caution that this covering fraction estimate may be rendered more uncertain by continuum 
variability and reverberation effects. However, the high-state continuum flux we observe is 
similar to the 1996-2003 historical average \citep{utm03}, mitigating these effects.
  
We treat C {\sc vi} separately from other ions in our model. The C {\sc vi} RRC 
is much stronger than expected, considering the apparently weak C {\sc vi}  2p-1s (Ly$\alpha$) 
emission line. This and the P-Cygni profile show that C {\sc vi} Ly$\alpha$ is heavily absorbed.
The C {\sc vi} emitter has a larger blueshift than the other ions, corresponding to 
$v=-310^{+70}_{-20}$ km s$^{-1}$. The emitter turbulent velocity width is much larger than 
the other ions ($b = 700 \pm 40$ km s$^{-1}$, $\mathrm{FWHM}=1200$ km s$^{-1}$), 
putting it in the BLR. The C {\sc vi} absorber also has a larger turbulent velocity 
($b = 410 \pm 20$ km s$^{-1}$) than other ions, but has the same blueshift. It covers a large 
fraction of the C {\sc vi} emitter, with $f_\mathrm{los}=0.88 \pm 0.02$. The narrow emission line 
component of C {\sc vi} Ly$\alpha$ may be seen at 33.8 \AA\ as a $~\sim 3\sigma$ residual to 
the fit (Fig. 5). All of this indicates that  C {\sc vi} emission arises primarily in the BLR. 
\cite{ksr02} find a similar (though much broader) C {\sc vi} Ly$\alpha$ line in NGC 5548 which 
they attribute to the BLR.

The  C {\sc vi} RRC is strong enough to fit for the BLR electron temperature which determines 
its width. We estimate $T_\mathrm{e}=4.0^{+3.3}_{-1.3}\times 10^4$ K, which is consistent with the value 
$3.2\times 10^4$ K predicted by XSTAR for a photoionized nebula with maximum C {\sc vi} abundance. 
The rest of the RRCs are too weak or blended to measure accurately.

In addition to narrow O {\sc vii} emission, there appears to be broad
emission at the location of the O {\sc vii} 2p-1s (r) line (Fig. 5). Adding a broad Gaussian 
to the fit improves it by a significant amount ($\Delta\chi^2=27$, for 3 additional degrees of
freedom). The best-fit parameters are $\lambda=21.7 \pm 0.1$ \AA, $\sigma=0.33 \pm 0.08$ \AA, 
and a flux of $2.2 \pm 0.7 \times 10^{-4}$ ph s$^{-1}$ cm$^{-2}$. This indicates a component
of O {\sc vii} emission from the BLR with FWHM$=11,000 \pm 3000$ km s$^{-1}$. Flux from
the intercombination lines may also contribute significantly, depending on the BLR density.

Finally, we note some residual absorption lines in the spectrum (Fig. 5). The line at 
16.8 \AA\ may be residual Fe {\sc xvii} absorption which cuts into the O {\sc vii} RRC. The 
line at 30.3 \AA\ is narrower than the instrumental response and appears only in RGS1, so 
is likely to be an unmarked bad column. There is also an unidentified absorption line at 
36.4 \AA.
    
\subsection{Low State}
    
The low-state RGS spectrum of NGC 4051 (Fig. 7) contains prominent narrow
O {\sc vii} emission lines on a weak continuum. There may be weak Fe L-shell 
UTA (unresolved transition array) features which contribute to a broad dip in the 
spectrum from 15-18 \AA. Alternatively, this may be a gap between blended narrow 
Fe L-shell emission lines. Because of the low flux level and short duration of this
particular low state, it is not possible to characterize and measure the spectral 
features with any great accuracy. NGC 4051 was more recently observed by XMM during 
an extended low state as a target of opportunity in 2002 November \citep{utm03, p03}.

We model the low-state spectrum with a small number of components, minimizing the 
C-statistic to determine the best fit to the data. Our best-fit model has a steep power 
law continuum with $\Gamma = 2.60$ and a normalization of $4.49 \pm 0.05 \times 10^{-5}$ 
ph s$^{-1}$ cm$^{-2}$ \AA $^{-1}$ at 1 \AA. The corresponding 0.3-10 keV luminosity is
$L_\mathrm{x}=1.3 \times 10^{41}$ erg s$^{-1}$. The soft power law continuum index is a bit 
steeper in the low state than in the high state. 

Unlike the high state, there is no obvious curvature or broad bump in the low-state 
soft X-ray spectrum. We tried adding the relativistic O {\sc viii} emission line series, with
all disk parameters fixed to high-state values, except for flux normalization. 
We find a (90\%) upper limit of EW$<150$ eV for the entire line series. This limit is only
slightly smaller than the EW observed during the high state (164 eV). We can neither confirm
nor rule out the presence of relativistic line emission during the low state. The EPIC spectra 
show a correlation between the relativistic O {\sc viii} emission and the flux of the soft 
X-ray continuum \citep{spw03}. However, it is difficult for EPIC to constrain the relativistic
O {\sc viii} line flux during the low state.

We include only the strongest absorption and emission line features in our low-state model.
The NLR and continuum covering fractions are fixed at 1.0, to match the high state.
We find an O {\sc vii} absorption column density of $1.6^{+0.8}_{-0.5} \times 10^{17}$ cm$^{-2}$ 
and a Doppler parameter of $b=230 \pm 30$ km s$^{-1}$ in the narrow absorber. Both are  
consistent with the parameters observed during the high state. The weak continuum and low S/N make 
it difficult to derive accurate column densities for the other K-shell ions. In order to quantify 
the amount of Fe UTA absorption, we assume equal column densities of Fe {\sc i}-{\sc xvi}. The best 
fit mean column density is $1.3 \pm 0.1 \times 10^{16}$ cm$^{-2}$ in each of these ions. The 
Fe {\sc xvii}-{\sc xx} L-shell transitions are too weak to measure.

Two ions in the NLR emitter with measurable column densities are O {\sc vii}
($4.0\pm 0.4 \times 10^{17}$ cm$^{-2}$) and O {\sc viii} ($4^{+2}_{-1}\times 
10^{16}$ cm$^{-2}$). The O {\sc vii} emitter column density appears to be a factor of 1.3 greater
during the low state than during the high state. However, this difference may be explained
by the poor resolution of the low state data, blending of lines in the O {\sc vii} 
triplet, or residual emission from the BLR rather than a true variation. The best-fit 0.3-10 keV 
normalization for the NLR emitter is 
$f_\mathrm{c} L_\mathrm{x}=4.4 \pm 0.3 \times 10^{41}$ erg s$^{-1}$ cm$^{-2}$, unchanged from the 
high state. 

The O {\sc vii} f narrow emission line flux is $8.9\pm 3.8 \times 10^{-5}$ ph s$^{-1}$ 
cm$^{-2}$ during the low state, compared to $8.8\pm 1.6 \times 10^{-5}$ ph s$^{-1}$ cm$^{-2}$ 
during the high state. There is no significant variability, which is consistent with the 
expected $r>0.02$ pc size scale of the NLR (\S 4.4). Dividing $f_\mathrm{c} L_\mathrm{x}$ from the narrow 
emission lines by $L_\mathrm{x}$ from the power law continuum gives an unphysical global covering 
fraction of $f_\mathrm{c}=3.5\pm 0.2$. This is an indication that the NLR does not respond to short 
timescale variations in the continuum flux, and that values of $f_\mathrm{c}$ derived this way are highly 
uncertain.

\subsection{Comparison to {\it XMM-Newton} EPIC Data}

The full EPIC data set and spectral variability are analyzed in detail by \cite{spw03}, and we 
do not repeat those results here. However we do test for consistency between the RGS and EPIC 
PN high-state data sets. The EPIC PN data were filtered to exclude the low state and the period of 
high background during the last 14 ks. We use IMP  to fit the 0.3-10 keV band with a model 
consisting of hard plus soft power laws, relativistically broadened O {\sc viii} and Fe {\sc xxv} 
K-shell recombination emission (line series + RRC), and two narrow Gaussians to represent O {\sc vii} and 
Fe K$\alpha$ emission lines. The resulting fit and residuals are shown in Fig. 8. The fit is 
remarkably good ($\chi^2/DF=1.39$), considering  the low resolution of the EPIC camera at soft 
energies and crudeness of the model relative to our REL2 fit of the RGS spectrum.

The EPIC data are fit by a hard power law with $\Gamma=1.52$ and a soft power law with $\Gamma=3.01$. 
The soft power law is considerably steeper than we find for the RGS data. This points to a 
cross-calibration problem\footnote{Calibration document available at 
\url{http://xmm.vilspa.esa.es/docs/documents/CAL-TN-0018-2-1.pdf}}
between the instruments which has been observed for other sources \citep{cal03}. Uncertainty in the 
PN response at low energy may be at fault.  The mean 2-10 keV EPIC flux for the entire observation, excluding 
the background flare, is $F_{2-10}=2.2 \times 10^{-11}$ erg s$^{-1}$ cm$^{-2}$, corresponding to a luminosity
of $L_{2-10}=2.3 \times 10^{41}$ erg s$^{-1}$. This is identical to the mean {\it RXTE} 2-10 keV flux for the 
years 1996-2003 \citep{utm03}, indicating that NGC 4051 was in a typical state. 

Narrow Fe K$\alpha$ has a rest energy of $6.41\pm 0.04$ keV, consistent with fluorescence from low charge 
states Fe {\sc i-xviii}. The (90\% confidence) upper limit to its width is $\sigma=0.17$ keV, so
it is unresolved by EPIC (FWHM$<18,000$ km s$^{-1}$). The ratio of the data to a power law with
Galactic absorption shows a broad excess in the 4-7 keV region, which may be identified with
relativistic Fe K$\alpha$ \citep{spw03}. The best fit Laor model has $R_\mathrm{in}<2.1 R_\mathrm{G}$ and
$i\sim 55\arcdeg$, roughly consistent with relativistic O {\sc viii} in the RGS spectrum. The 
relativistic Fe K$\alpha$ line flux is $2.1 \times 10^{-8}$ ph s$^{-1}$ cm$^{-2}$. The rest energy is 
$E=6.70^{+0.01}_{-0.03}$ keV, indicating that Fe {\sc xxv} is the dominant ionization stage.

We find significant emission from relativistic O {\sc viii}, seen at $\sim 0.6$ keV in 
a plot of the ratio of data to the 2 PL continuum (Fig. 8c). It appears more sharply peaked than in 
the RGS spectrum because of blending with the O {\sc vii} narrow line. The best fit Laor model 
has  $R_\mathrm{in}<1.6 R_\mathrm{G}$ and $i = 59 \pm 8 \arcdeg$ \citep{spw03}.

\subsection{Comparison to {\it Chandra} HETGS Data}

As discussed above, the broad bump in the NGC 4051 soft X-ray excess has been noted
by previous authors, and typically fit by a single-temperature blackbody. Here we test our
new interpretation, that this excess is from relativistically broadened O {\sc viii} emission,
against the 2000 April {\it Chandra} HETGS observation. The soft X-ray bump is apparent during
this epoch (Fig. 9), and has a similar shape to the bump in the 2001 May {\it XMM-Newton} RGS data.
We fit the  1.7-22.6 \AA\  {\it Chandra} MEG spectrum with IMP, using $\chi^2$ to assess goodness of 
fit. The model is similar to REL2, but we add a hard X-ray power law to account for the upturn in 
the spectrum below 5 \AA. We also add Ne {\sc X}, Mg {\sc xi-xii} emission and absorption to the model, 
and remove the lower ionization states C {\sc v-vi}, N {\sc ii-vi}, and O {\sc iv-vi} which are either 
too weak to model or are outside the MEG waveband. 

The best fit REL model has $\chi^2/DF=1.59$. One source of residuals may be the effective area calibration 
uncertainty at the Ir and Au instrumental edges at 5-6 \AA. There appear to be additional residuals in the 
8-14 \AA\ region from Fe L-shell absorption and emission lines. There is another interesting residual
at the location of the narrow O {\sc viii} Ly$\alpha$ line, which appears to be stronger than in the 
{\it XMM-Newton} observation.

The underlying continuum is fit by the sum of two power laws, a soft power law with $\Gamma=2.24$
and a hard power law with $\Gamma=1.30$. The soft slope is similar to that observed with
{\it XMM-Newton} RGS, while the hard slope is 0.2 dex harder than that observed with EPIC in 2001 May. 
The normalizations are $A_\lambda=1.51\pm 0.03 \times 10^{-4}$ ph s$^{-1}$ cm$^{-2}$ \AA$^{-1}$ for 
the soft power law and $A_\lambda=6.9\pm 0.2 \times 10^{-4}$ ph s$^{-1}$ cm$^{-2}$ \AA$^{-1}$ for the 
hard power law, both evaluated at 1 \AA. The soft X-ray flux at 12.4 \AA\ (1 keV) is a factor of 
$\sim 2$ lower in the {\it Chandra} MEG spectrum than in the {\it XMM-Newton} RGS spectrum.

The soft X-ray excess is fit with a relativistically broadened O {\sc viii} line series and RRC.
We assume emission from a Keplerian disk around a Kerr black hole, producing \cite{l91} line
profiles. The inner radius is fixed at $R_\mathrm{i}=1.24 R_\mathrm{G}$ and the outer radius is fixed 
at $R_\mathrm{o}=400 R_\mathrm{G}$. The best-fit disk inclination is $50 \pm 1\arcdeg$. The best-fit 
index for the radial power law emissivity profile is $q=5.9 \pm 0.2$, steeper than for the {\it XMM-Newton} 
RGS spectrum.  If we take into account the positive correlation between $i$ and  $q$ (Fig. 4c), and the lack 
of long-wavelength coverage of the MEG, we find that the {\it Chandra} and {\it XMM-Newton} results are 
roughly consistent. A more detailed comparison is hampered by the lower S/N in the MEG data at long wavelengths.

We conclude that a relativistic O {\sc viii} emission model gives a good fit to the NGC 4051 
spectrum for two separate epochs, viewed with two different X-ray observatories. This argues
against instrumental effects as the source of the unusual continuum shape. In addition, a secondary 
soft X-ray bump from 13-17 \AA\ is clearly visible in both the {\it Chandra} (Fig. 8) and 
{\it XMM-Newton} (Fig. 2) data. This detection of high-order emission from the O {\sc viii} line series 
reinforces our interpretation of relativistic line emission from the photoionized inner disk.

\section {Discussion}

\subsection{Basic AGN Parameters}

NGC 4051 becomes the third known Seyfert galaxy with strong soft X-ray relativistic 
lines (SRLs). As noted by \cite{br01}, these all belong to the class of narrow-line 
Seyfert 1 (NLS1) galaxies. NLS1's Ton S180 \citep{tgy01}, Ark 564 \citep{vpr99}, and IRAS 
13224-3809 \citep{btf03} also display a broad, curved soft excess which may include relativistic 
emission lines. In addition, these galaxies show evidence for ionized reflection in the hard 
X-ray band \citep{bif01,btf03}. Weaker (EW$\sim$20 eV) SRL features may be present in the
spectra of broad-lined Seyfert 1's NGC 5548, NGC 4593, and MCG-2-58-22 \citep{ksr02,myg03,salv03a}.

NLS1's share a number of unique properties, including 1) narrow hydrogen Balmer emission 
lines ($v<2500$ km s$^{-1}$) \citep{g89}, 2) highly variable X-ray spectrum \citep{l99a}, and 
3) steep soft X-ray spectral index \citep{bbf96}. These properties may be explained if NLS1's 
contain lower mass black holes which accrete at higher fractions of the Eddington accretion rate 
than other Seyfert 1 galaxies. Note that MCG-6-30-15 is a borderline NLS1, with  
FWHM(H$\beta$) = 2500 km s$^{-1}$ \citep{rwf97} and moderate ($\Gamma=1.8$) soft X-ray spectral 
index. However, it tends to be included in this class due to its highly variable X-ray continuum 
and inferred black hole mass of $\sim 10^6 M_\odot$ \citep{nc00}.

\cite{p00} obtain a reverberation estimate of the mass of the black hole in NGC 4051 of 
$M = 1.1^{+0.8}_{-0.5} \times 10^6 M_{\odot}$, roughly a factor of 10 lower than the mass for 
broad-lined Seyfert 1s with similar luminosities. A similar estimate by \cite{sun03} gives
$M = 5^{+6}_{-3} \times 10^5 M_{\odot}$, assuming isotropic circular orbits for the BLR clouds. 
We adopt the latter, slightly more accurate value for our subsequent analysis. 
Reverberation mass estimates are subject to uncertainty in the BLR geometry and kinematics \citep{k01}. 
The largest uncertainty comes in the case of a thin disk, where projected velocity width goes as 
$\sin i$ and the virial mass may be severely underestimated for small inclination. However,
if the large RLR inclination of $48\arcdeg$ also applies to a disk-like BLR, the virial mass
is {\it reduced} by the factor $ 1 / 3 \sin^2 i = 0.6$. This reinforces the interpretation that 
NGC 4051 and NLS1s in general contain relatively low mass black holes. The high inclination can be 
used to argue against a pole-on orientation as the cause of the narrow permitted lines from the BLR.

From the black hole mass, we compute the Eddington luminosity $L_\mathrm{E} \sim 6 \times 10^{43}$
erg s$^{-1}$. The observed high-state bolometric luminosity is $\sim 50\%$ of the Eddington value. 
The accretion rate is $\dot{M} = 4.7 \times 10^{-4} \eta^{-1} M_\odot$ yr$^{-1}$, where
$\eta$ is the efficiency of the accretion process. Assuming a typical value of $\eta=0.1$, the 
mass doubling time is $\tau = \ln (2) M/\dot{M} \sim 7 \times 10^7$ yr. At this rate, the AGN could 
plausibly have originated from a $10 M_\odot$ black hole near the center of NGC 4051 $\sim 1$ Gyr ago.

\subsection{Relativistic Line Region}

With 3 objects in the class of strong SRL emitters, we can begin to compare their properties. 
MCG-6-30-15 and Mrk 766 have similar disk inclinations, emissivity indices, 
and relativistic N {\sc vii}/O {\sc viii} emission line ratios \citep{br01}. NGC 4051 stands 
out for its much broader lines and absence of relativistic N {\sc vii} emission. It has a steeper
($\Gamma=2.4$) soft X-ray continuum, compared to $\Gamma=1.8, 2.1$ in MCG-6-30-15 and 
Mrk 766 \citep{skb03,mbo03}. Of the three objects, NGC 4051 also has the greatest excess variance 
in its X-ray light curve on short (10-100 ks) timescales \citep{ngm97a}. The C {\sc vi}/O {\sc viii} 
emission line ratio ranges from 0.6 in Mrk 766, to 0.15 in MCG-6-30-15 \citep{br01}, to $<0.14$ in NGC 4051. 
The lack of relativistic C {\sc vi} Ly$\alpha$ emission may indicate that the inner disk is more highly ionized 
in NGC 4051 than in MCG-6-30-15 and Mrk 766. This is consistent with a smaller inner disk radius and 
stronger, steeper, more highly variable soft X-ray continuum.

The strong SRL emitters are parameterized by steep radial emissivity laws, with 
most of the emission coming from very close to the central black hole. The rapid 
X-ray continuum variability in NGC 4051 and Mrk 766 is correlated with the relativistic 
O {\sc viii} line variability \citep{spw03,mbo03}, perhaps indicating that the continuum drives
the line. (A similar variability analysis has yet to be performed for MCG-6-30-15.)  Bright flares 
(with amplitude $>100\%$) in the X-ray continuum occur on a  timescale of 3 ks \citep{mmp02}, yielding 
an upper limit of 50 light minutes ($600 R_\mathrm{G}$) for the size of the flaring regions. 

All strong SRL emitters also have relativistic Fe K$\alpha$ lines. Some (or all) of this emission may come 
from Fe {\sc xxv} and Fe {\sc xxvi} \citep{mbo03,bvf03,spw03} in a photoionized disk. This is especially 
important for understanding the broad Fe K line in MCG-6-30-15, where so much effort has gone into 
understanding this emission in terms of Fe {\sc i} fluorescence. The problems with $>100\%$ reflection 
fraction and iron overabundance may disappear if most of the line emission comes from recombination in an 
ionized disk \citep{bvf03}.

To characterize the RLR and other emission line regions, we compute XSTAR \citep{km82, k02} 
photoionization models for an optically thin plasma (using the SED in Fig. 2). The illumination 
of the NLR and BLR is probably well approximated by an $r^{-2}$ drop in flux with radius $r$ from
a central point source. The geometry must be more complicated for the RLR, but the central point 
source assumption is useful for an order of magnitude ionization estimate. In Figure 10, we plot 
ionization parameter $\xi=L/n_\mathrm{e} r^2$ vs. line FWHM. Line width is converted
to radius $r$ assuming isotropic circular Keplerian orbits around a central black hole with mass 
$5\times 10^5 M_{\odot}$. For this mass, the innermost circular orbit around a Kerr black hole is 
at $3 \times 10^{-8}$ pc (3 light-seconds). The $r$-$\xi$ parameter space of the ionization map
is populated by the emission lines observed in the X-ray and optical-UV spectra, assuming they are 
produced in regions of peak emissivity. In this way, we estimate the densities needed to explain 
the ionization levels observed in the RLR, BLR, and NLR. 

The relativistic O {\sc viii} emission line profile is produced near the inner edge of the accretion
disk at $\sim 1.6 R_\mathrm{G}$ from the black hole. Though the emissivity of O {\sc viii} Ly$\alpha$ 
peaks at $\log \xi=1.7$, there would be strong C {\sc vi} and N {\sc vii} Ly$\alpha$ if the ionization 
parameter of the RLR were this low. The lack of relativistic recombination emission from C {\sc vi} gives
a lower limit of $\log \xi>3.4$ (Fig. 11). Fe {\sc xxv} emission peaks at an ionization parameter of 
$\log \xi=3.4$ and may contribute to the relativistic Fe K line. The photon flux ratio of relativistic 
Fe K$\alpha$ (measured with EPIC) to O {\sc viii} Ly$\alpha$ is 0.06. This is close to the peak 
ratio of Fe {\sc xxv}/O {\sc viii} Ly$\alpha$ $\sim 0.08$  predicted by our XSTAR photoionization model, 
assuming solar abundances.

Assuming the mean distance from the X-ray source to the surface of the photoionized disk is 
$1.6 R_\mathrm{G}$, $\log \xi>3.4$ implies a density of $n_\mathrm{e} < 1 \times 10^{17}$ cm$^{-3}$ in 
the RLR. This is only changed by a factor of order unity if we assume a geometry where the disk 
intercepts half of the flux from an X-ray emitting corona. This upper limit falls between densities 
predicted for radiation dominated ($n_e=10^{13}\alpha^{-1}$ cm$^{-3}$) and gas-pressure dominated disks 
($n_e=10^{20}\alpha^{-7/10}$ cm$^{-3}$), where $\alpha\sim0.1$ provides the scaling between pressure and 
viscous stresses in the disk \citep{ss73}. However, these are only rough estimates, and do not take into 
account general relativistic effects on the disk structure around a Kerr hole \citep{nt73}.

N {\sc vii} and O {\sc viii} should coexist at very similar ionization parameters and densities
in the inner disk. However, for solar abundances, the nitrogen to oxygen ratio is 0.13, 
explaining the lack of relativistic N {\sc vii} emission in the spectrum of NGC 4051. It is 
actually more puzzling that there is strong relativistic N {\sc vii} emission in MCG-6-30-15 and 
Mrk 766 \citep{br01}, which requires nitrogen overabundances of 6-9 times solar. Alternatively, 
N {\sc vii} emission could be pumped by O {\sc viii} Ly$\alpha$ via the Bowen 
resonance-fluorescence mechanism \citep{s03}. If this is the case, it may indicate a larger 
N {\sc vii} optical depth in these two sources than in NGC 4051. 

Assuming $\log \xi=3.4$ in the inner disk, the fraction of oxygen in O {\sc viii} is 
$5.1 \times 10^{-3}$. (The rest is in O {\sc ix}.) The O {\sc viii} edge has an optical depth 
of $\tau \sim 1.0$ for the O {\sc viii} column density of $\sim10^{19}$ cm$^{-2}$ inferred from 
the relativistic O {\sc viii} Ly$\alpha$ EW. For a density of $n_\mathrm{e}= 1 \times 10^{17}$ cm$^{-3}$,
this is reached in a thickness of 230 km. For solar oxygen abundance, the Thomson optical depth is 1.5, 
indicating that electron scattering is the dominant source of opacity in the O {\sc viii} 
Ly$\alpha$ emitting layer. Compton scattering is likely to cause additional line broadening 
\citep{brf02}, which has not been accounted for in our model. The $2 \times 10^6$ K temperature
of a layer at $\log \xi=3.4$ is considerably cooler than the Compton temperature 
($\sim 1 \times 10^7$ K).

At the high temperatures and densities inferred for the RLR, there may be a non-negligible 
contribution from bremsstrahlung to the observed soft X-ray continuum. For 
$T_\mathrm{e}=2.0\times 10^6$ K, a bremsstrahlung spectrum cuts off above 0.17 keV, but the exponential 
tail is potentially observable at the long wavelength end of the RGS band. The emission measure 
of the  relativistic O {\sc viii} emission region is $n_\mathrm{e}^2 V=8.2\times10^{64}$ cm$^{-3}$ from 
our REL2 model. Assuming solar abundances and  $\log \xi=3.4$, bremsstrahlung from the RLR could 
contribute as much as $18\%$ of the total flux at 38 \AA. We see no evidence of this in the RGS spectrum,
but it can not be excluded. Coverage at longer wavelengths would be useful to put better constraints on 
this component of the continuum emission.

The detection of higher-order relativistic O {\sc viii} emission in NGC 4051 indicates moderate optical 
depth in the photoionized layer of the accretion disk. It also begs the question of whether or not these 
lines are present in MCG-6-30-15 and Mrk 766. The sharper line profiles in the latter two objects would 
lend themselves to detection of the O {\sc viii} Ly$\beta$ line. However, this region of their spectra 
is also highly absorbed by Fe UTA features. The absence of Fe UTAs in the high-state spectrum of NGC 4051 
yields a simpler model and easier detection. \cite{br01} identify a sharp feature at 
15-16 \AA\ in MCG-6-30-15 which could be O {\sc viii} Ly$\beta$ from the RLR. However, they 
argue instead that it may be an O {\sc viii} absorption edge. This issue should be revisited
in light of our current results, and may have important implications for the optical depth of 
the O {\sc viii} RLR in these sources. 

By extension, we predict emission in Ly$\beta$ and higher order lines from relativistic Fe {\sc xxv} 
and Fe {\sc xxvi} in the ionized accretion disk. It will be important to test this 
hypothesis against higher resolution data from {\it Astro-E} and {\it Constellation-X}. This 
effect may help to distinguish between fluorescence and recombination models of the Fe K$\alpha$ 
emission line in Seyfert galaxies. As we find for the O {\sc viii} series, including higher order
lines in the spectral model may significantly change the accretion disk fit parameters. It will 
also be important to consider this emission when studying absorption features in the Fe K-shell 
region \citep[e.g.]{prp03}. 

We would like to stress that while emission from a Keplerian disk inside a Kerr spacetime
geometry is consistent with the observed O {\sc viii} profile, it is not the only 
explanation. Alternatively, this emission may come from inside the innermost stable 
circular orbit of a Schwarzschild black hole \citep{rb97}. Also, the parameterization of 
the radial emissivity by a power law is not necessarily correct. Other emissivity laws which
peak sharply inside of $1.6 R_\mathrm{G}$ could produce a similar line profile. However, it is 
unlikely that the observed line profile comes from a two-sided collimated jet or a relativistic 
wind \citep{fnr95}. 

\subsection{Broad Line Region}

The broad C {\sc vi} Ly$\alpha$ emission line can be produced in the BLR of NGC 4051 at a 
velocity of $1200$ km s$^{-1}$ with $n_\mathrm{e}=1\times 10^{10}$ cm$^{-3}$ at roughly 
$2 \times 10^{-3}$ pc (2 light-days) from the central source (Fig. 10). The core of the C {\sc iv} 
broad line is produced optimally at a density of $n_\mathrm{e}=2\times 10^{11}$ cm$^{-3}$. The much lower
ionization H$\beta$ line has a similar velocity width (FWHM$ = 1100$ km s$^{-1}$) and 
reverberation size of $3\pm 1.5$ light days \citep{p00,sun03}, and would be emitted optimally 
at a density of  $n_\mathrm{e}=4\times 10^{15}$ cm$^{-3}$. This requires a density contrast of 
$4\times 10^5$ for all three lines to be produced in the same region.  The net blueshift of C{\sc vi}
Ly$\alpha$ with respect to the galaxy rest frame and P-Cygni profile may indicate outflow in the 
outer BLR.

In addition to their line cores, the He {\sc ii} $\lambda 4686$ and C {\sc iv} $\lambda 1549$ lines 
in NGC 4051 have very broad blue wings  with FWHM$=5400$  km s$^{-1}$ \citep{p00}. It is likely that
the high velocity component is located closer to the central black hole than the low-velocity 
component. This is supported by the large RMS variability observed in broad  He {\sc ii} \citep{p00}. 
The high velocity He {\sc ii} and C {\sc iv} lines are then produced at high densities 
($n_\mathrm{e}=2\times 10^{14}-2 \times 10^{16}$ cm$^{-3}$), roughly $9 \times 10^{-5}$ pc from the 
center (Fig. 10). Very broad O {\sc vii} is produced at a density of $n_\mathrm{e}=7 \times 10^{13}$ at 
$2 \times 10^{-5}$ pc, four times closer to the central black hole. Broad (FWHM $\sim 5000$ km s$^{-1}$),
blue-shifted components of C {\sc iv} and H-Ly$\alpha$ are common to a number of NLS1s \citep{rp97}, and 
have been attributed to an outflowing BLR \citep{l00}.

With the structure of the line emitting regions shown in Fig. 10, we are led
to ask why certain regions of parameter space are filled and others are not. The outer radius 
of the BLR may be set by the dust sublimation radius \citep{nl93,l03}, with $R_\mathrm{sub}\simeq 
0.2(L_\mathrm{bol}/10^{46})^{1/2}$ pc. For NGC 4051, $R_\mathrm{sub}\simeq 1.0\times 10^{-2}$ pc, which 
is close to the estimated BLR outer radius in Fig. 10. The inner radius may be constrained to be greater 
than the ionizing UV emission region. For a disk blackbody emitting at 50\% of the Eddington limit, 
$R_\mathrm{BLR}>R_\mathrm{UV}\sim 200 R_\mathrm{G}$ ($5\times10^{-6}$ pc). The lower right hand corner 
of Fig. 10 is empty, indicating that the highest density emission line regions have 
$n_\mathrm{e}<10^{17} $ cm$^{-3}$. The accretion disk density may set this upper limit for 
the BLR density, especially if it is the source of the BLR plasma. The lack of emission lines from 
the BLR with ionization greater than C {\sc vi} may owe to insufficient S/N.

Observations are qualitatively consistent with a model where the BLR is formed as an 
accretion disk wind \citep{mc97}. In this scenario, gas from the disk is heated and rises to a 
location where it is exposed to and accelerated by UV photons. The velocity width of 
the broad lines is produced primarily by Keplerian rotation, but radial velocity shear in the 
wind yields a single-peaked profile. The broad base of the line comes from  close in while the 
narrow core arises further from the black hole. An accelerating outflow would naturally span a large 
range of densities at any given radius, from the density of the neutral disk surface to the much 
lower density, highly ionized C {\sc vi} emitting region observed with {\it XMM-Newton}. The BLR 
appears to span the entire range of allowed radii, from just outside the UV continuum emission region
to where it merges (perhaps seamlessly) with the NLR. In the case of NLS1s such as NGC 4051,
the distinction between the BLR and NLR may be artificial.

\subsection{Narrow Line Region}

The narrow N {\sc vi}, O {\sc vii}, Ne {\sc ix}, and  Si {\sc xiii} emission lines require 
densities of $n_\mathrm{e}=10^7-10^8$ cm$^{-3}$ at $r>0.02$ pc (Fig. 10). These are below the critical 
densities of the forbidden lines ($\log n_\mathrm{crit} \sim 9-13$) and consistent with 
the observed ($f/i>3$) line ratios \citep{pd00}. A low-state {\it Chandra} ACIS-S image of 
NGC 4051 shows little extended emission \citep{ufh03}. This is different from other nearby 
Seyferts, including NGC 4151, Circinus, and NGC 1068 \citep{o00,snk01,obc03} where the X-ray 
NLR is extended on the same size scale as the optical NLR. 

The optical [O {\sc iii}] emission region has a smaller velocity width 
(FWHM $=240$ km s$^{-1}$) than O {\sc vii} (Fig. 10), and extends to a maximum distance of 220 pc 
from the nucleus \citep{chs97}. (The contribution of the galactic stellar bulge to the virial
mass dominates at these radii, leading to velocities of a couple hundred km s$^{-1}$.) The 
[O {\sc iii}] $\lambda 5007$ line has a critical density of $n_\mathrm{e} \sim 10^6$ cm$^{-3}$, which
puts a lower limit of $\sim 30$ pc (0\farcs 7) on its inner emission radius. This implies that
there may be a separation between the X-ray and optical narrow line regions. 

The large fractional covering of the X-ray NLR by the ionized X-ray absorber requires the absorber
to be at larger radius than the emitter ($r>0.02$ pc). The conditions in the narrow absorber are 
likely to be similar to the emitter, but may have somewhat lower densities to attain the same 
ionization parameter at larger radius. If we assume they have the same density and radius, then the 
column densities in Table 1 give a radial thickness of $\sim 10^{13}$ cm, and radial filling 
factor of $\sim 10^{-4}$ for the ionized absorber-emitter. This small value, plus the large 
global covering fraction indicate a sheet-like morphology. 

The accretion disk inclination implies a lower limit on the UV-X ionization cone opening 
half-angle of $\theta \ge 48\arcdeg$. This is larger than the value of $\theta=23\arcdeg$ for the 
[O {\sc iii}] conical outflow model \citep{chs97}. It appears that the [O {\sc iii}] cone is
narrower than the full ionization cone. The O {\sc vii} global covering fraction 
yields a half-opening angle of $77\arcdeg$ for the X-ray ionization cone, which is larger than 
but consistent with the disk inclination angle. However, variability could cause us to 
overestimate the covering fraction and ionization cone opening angle.

We estimate the outflow rate of the O {\sc vii} NLR, assuming it resides in a 
thin shell of covering fraction $f_\mathrm{c}$. Combining the equation of continuity with the definition
of the ionization parameter $\xi$, the mass outflow rate is 

\begin{equation}
{\dot M}_\mathrm{out} = 4 \pi f_\mathrm{c} {L_\mathrm{ion} \over \xi} \mu m_\mathrm{p} v_r,
\end{equation}

\noindent
where $\mu=1.3$ is the mean atomic weight per electron and $v_r=400$ km s$^{-1}$ is the observed 
radial outflow velocity of the absorber. Using $\log \xi =1.0$ from our XSTAR simulation,
$f_\mathrm{c}=0.76$ from the O {\sc vii} f equivalent width, and 
$L_\mathrm{ion}=4.1 \times 10^{42}$ erg s$^{-1}$, we find a large mass outflow rate of 
$\sim 5 M_{\odot}$ yr$^{-1}$. This is a robust result since we have directly measured most of the 
variables in Equation 1. The quantity $f_\mathrm{c} L_\mathrm{ion}$ comes directly from the emission line 
normalization, averaged over the entire emission region. The absorber radial 
velocity is probably a fair estimate of the mean flow velocity.

The estimated mass outflow rate is a factor of $10^3$ greater than the nuclear accretion 
rate, assuming an efficiency of $\eta=0.1$ (\S 4.1). The kinetic luminosity of the outflow 
is a significant fraction ($\sim 7\%$) of the ionizing luminosity of the AGN, but only a small
fraction ($\sim 1\%$) of the bolometric luminosity. A lower (but still suprisingly large) mass 
outflow rate of $3 M_{\odot}$ yr$^{-1}$ (10 times the accretion rate) is found for the O {\sc vii} 
NLR of the broad-lined Seyfert 1 galaxy NGC 3783 \citep{brb03}. If AGN outflows are driven by radiation 
pressure, then there may be a connection between the near-Eddington luminosity of NGC 4051 and its
large outflow rate. If this wind escapes the black hole, then the supply of fuel to the 
AGN will be exhausted more rapidly by outflow than by accretion. 

A number of intrinsic UV absorption systems are observed by {\it HST}, with 
velocities ranging from $+30$ to $-647$ km s$^{-1}$, in ions Si {\sc ii}-{\sc iv}, 
C {\sc ii}-{\sc iv}, and N {\sc v} \citep{cbk01}. The UV absorber components 1-4 are 
the closest in velocity (-647 to -337 km s$^{-1}$) to the X-ray absorbers we observe 
with {\it XMM-Newton}. The N {\sc v} $\lambda 1239,1243$ lines appear to be saturated and 
\cite{cbk01} do not give column density estimates for them. We predict a mean optical depth of 
the N {\sc v} $\lambda 1239$ line of $\tau=2.8$ from the N {\sc v} column density (Table 1) and 
the UV line width of 730 km s$^{-1}$. Therefore we expect the UV lines to be nearly saturated, 
based on our X-ray observations. A similar result is found for NGC 5548 by \cite{aks03}, who compare
{\it XMM-Newton} and {\it FUSE} observations of  O {\sc vi} lines.

There is a strong trend in the ionized absorber-emitter of increasing column density with 
ionization parameter (Fig. 6). A linear least-squares fit to a power law 
$N_\mathrm{H}=N_\mathrm{A} \xi^\alpha$ (including only absorber $N_\mathrm{H}$) gives an index of 
$\alpha=d \log N_\mathrm{H}/ d \log \xi=0.5\pm0.2$ and intercept of $\log N_\mathrm{A} = 20.0\pm 0.3$.  
A similar trend is found in NGC 5548 by \cite{skv03}, with slope in the range 0.25 to 0.5. However,
a power law gives a formally poor fit to the absorber column density distribution in NGC 4051. 
The distribution appears to peak at $\log \xi \sim 1.4$, with the column density dropping off at
higher ionization parameter. 

The ionization distribution of the absorber may either be continuous or consist of 
discrete phases. The 3.7 decades of ionization parameter observed in the NGC 4051 outflow (Fig. 6)
require at least 2 phases. Two-phase models have been suggested as a way to confine photoionized clouds 
\citep[e.g.]{ke03}. More than one phase may exist at pressures where the heating curve is multi-valued 
\citep{kk01}. We construct the heating curve for NGC 4051 (Fig. 12) using XSTAR and the SED (Fig. 2).
We place absorber ions at their ionization parameters of peak abundance and find that not all  
ionization phases can be in pressure equilibrium. While there do exist segments of vertical slope
and marginal stability on the curve, these only apply for the high ionization species 
N {\sc vii}-Fe {\sc xx}. The pressure of the lowest ionization phase of the absorber is as much as 100 
times greater than the pressure of the high ionization phases (Fig. 12).

It is a common property of Seyfert galaxies that the lower ionization UV and X-ray absorbers have lower 
column densities than their high ionization X-ray counterparts. While UV column density measurements 
are in some cases plagued by saturation and partial covering effects, this is less of a problem for
the lower optical depth K-shell absorption lines in the X-ray spectrum \citep{aks03}. In the case of 
NGC 4051, the N {\sc vii} absorber carries 100 times as much mass as the O {\sc iv} absorber. This must 
be telling us something fundamental about how the NLR clouds are produced. Comparing Figs. 6 and 12, it 
appears that the highest column density ions, N {\sc vii}-Fe {\sc xx}, may reside in regions of marginal 
stability, perhaps confirming the prediction of \cite{kk01}. However, the lower ionization species must 
exist out of pressure equilibrium and may be rapidly destroyed by expansion and photoionization heating. 
This may potentially explain their lower column density. Gas pressure confinement is not necessary in a 
dynamic outflow which is continually replenished by the accretion disk. 

\section{Summary}

{\it XMM-Newton} RGS observations of NGC 4051 demonstrate the capability of high-resolution X-ray 
spectroscopy to uncover the nature of AGN accretion and outflows at scales ranging from just outside 
the black hole event horizon to the narrow line region. The soft X-ray excess emission is a combination 
of continuum and relativistic emission line components. It is incorrect to fit this with 
simple models such as a power law or blackbody. It is important to get the continuum model correct 
for fitting the spectra of ionized absorbers. Conversely, it is essential to account for
ionized absorption in order to accurately measure the underlying continuum and 
relativistic emission lines.

\subsection{Relativistic Emission Lines}
    
We successfully model the soft X-ray excess in the spectrum of NGC 4051 as 
a power law plus relativistic O {\sc viii} emission, including Ly$\alpha$, higher-order emission lines,
and recombination continuum. The line profiles require emission from inside the last 
stable orbit of a Schwarzschild black hole, or emission close to the last stable orbit of a 
Kerr black hole. The relativistic emission line region has an inner radius of $<1.7 R_\mathrm{G}$,
which may indicate a rapidly spinning black hole. The accretion disk is viewed at a relatively high 
inclination of $48 \arcdeg$. Together with the reverberation mass of $5 \times 10^5 M_\odot$, this 
confirms that NGC 4051 contains a relatively low-mass black hole. 

The spectrum of NGC 4051 is quite different from those of MCG-6-30-15 and
Mrk 766. The relativistic O {\sc viii} line in NGC 4051 is considerably
broader, and there is no indication of relativistic N {\sc vii} or C {\sc vi}
emission lines. The lack of  relativistic C {\sc vi} and N {\sc vii} emission is 
consistent with solar abundances and a high ionization parameter ($\log \xi >3.4$).
This contrasts with  MCG-6-30-15 and Mrk 766, which require super-solar abundances
or resonance fluorescence effects to explain strong N {\sc vii} emission.
Photoionization modeling indicates that the relativistic line emission comes
from the thin, high density ($n \sim 10^{17}$ cm$^{-3}$) photoionized skin of the 
accretion disk. The $\sim 90$ eV equivalent width of  O {\sc viii} Ly$\alpha$ can consistently
be produced in the this layer, with an additional contribution from higher order Lyman transitions. 

\subsection{Ionized Absorber-Emitter}

We fit an ionized absorber-emitter model for several 
ions in the narrow-line spectrum of NGC 4051. The ions cover nearly 4 decades of 
ionization parameter. The column density increases with ionization 
parameter, with high ionization states carrying more mass than lower
ionization states. The absorber and emitter have similar ionic column densities, 
and the X-ray NLR is covered by the absorber, indicating that the two 
arise in the same region. The large global covering fraction and high velocity of the
X-ray NLR imply a mass outflow rate which is $10^3$ times greater than the accretion rate.
An analysis of the heating curve in the photoionized absorber 
shows that the low ionization UV-X absorber can not be in pressure equilibrium with
the higher ionization phases of the X-ray absorber.

We find evidence for Doppler-broadened C {\sc vi} and O {\sc vii} emission lines in the X-ray
spectrum. The C {\sc vi} emission line has a velocity width of 1200 km s$^{-1}$, similar to  
the UV and optical broad lines. This may indicate a high ionization component within the BLR of 
NGC 4051. The implied large range of densities and observed P-Cygni line profiles are 
consistent with the BLR arising in an accretion disk wind. A very broad (FWHM $\sim 11,000$ km 
s$^{-1}$) O {\sc vii} line may arise in the innermost regions of the BLR, just 
exterior to the UV emitting portion of the accretion disk. 

\subsection{Low-state Spectrum}

Observation of NGC 4051 during a low X-ray flux state shows high equivalent width narrow emission lines.
The X-ray NLR does not respond to rapid X-ray continuum variations, consistent with its estimated 0.02 
pc size scale. However, we find evidence for low ionization Fe UTA absorption during the low-state, 
which is not present during the high state. The source is too weak during the low state to measure the 
relativistic O {\sc viii} feature. 
 
\acknowledgements

Thanks to Omer Blaes, Shane Davis, and Aristotle Socrates for insightful discussions
about accretion disks. Thanks to Ski Antonucci for many helpful comments which improved
the quality of the manuscript. This research was accomplished with the {\it XMM-Newton} Observatory,
an ESA science mission with instruments and contributions directly funded by ESA Member 
States and the USA. We have made use of the NASA/IPAC Extragalactic Database (NED) which is 
operated by the Jet Propulsion Laboratory, California Institute of Technology, under contract with 
the National Aeronautics and Space Administration. PMO was partly funded by NASA grant NAG-12390 and 
a National Research Council associateship. NJS acknowledges receipt of a PPARC studentship.

\vfill
\eject

\begin{figure}[ht]
\plotone{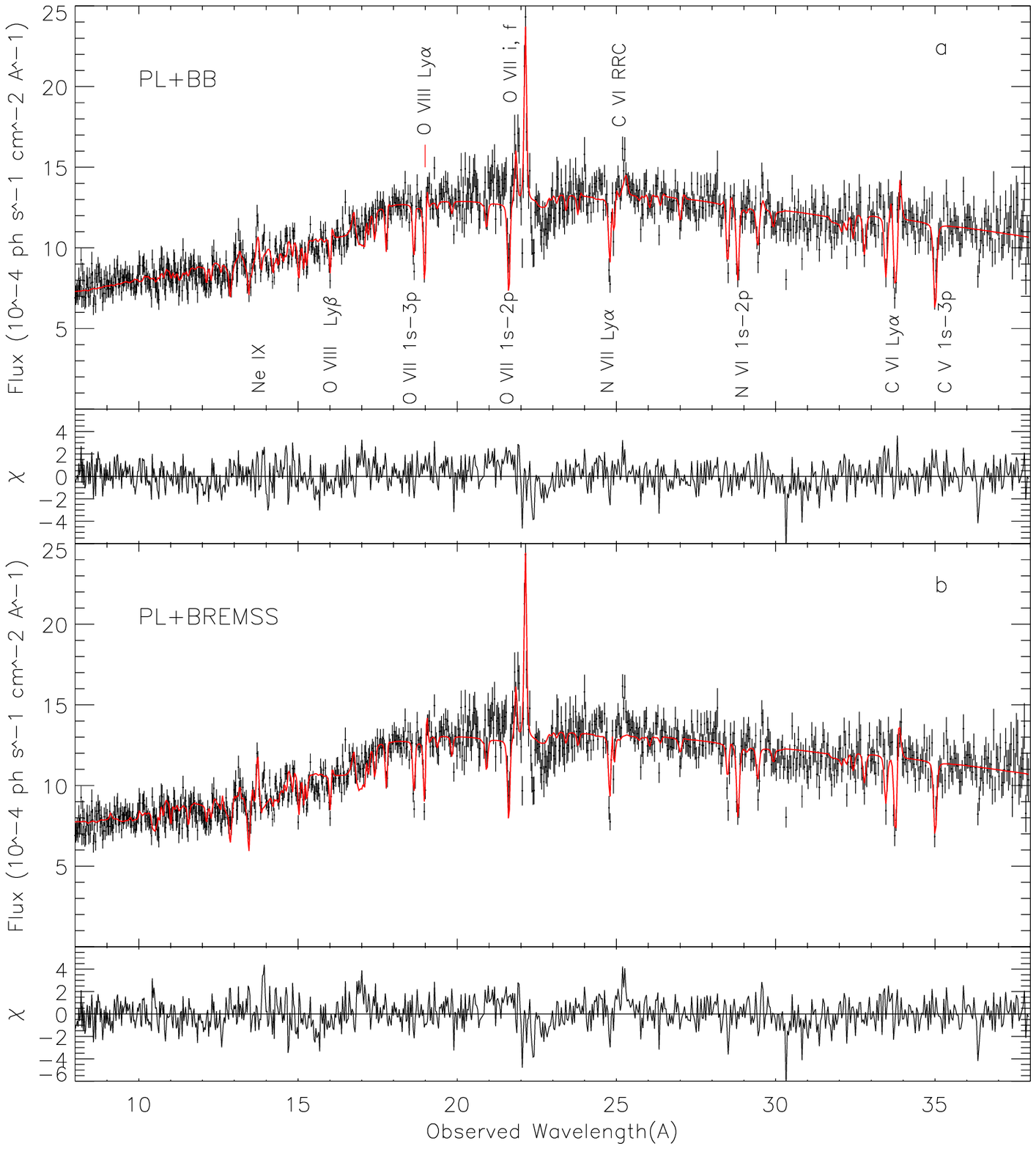}
\figcaption{Continuum model fits to the {\it XMM-Newton} RGS spectrum of NGC 4051. Narrow 
    absorption and emission lines are also included for various ions of C, N, O, Ne, 
    and Fe. a) Power law plus single blackbody. b) Power law plus bremsstrahlung.
    These are both poor fits, and fail to match the observed soft excess without 
    overestimating the O {\sc vii} edge depth.}
\end{figure}

\begin{figure}[ht]
\epsscale{1.0}
\plotone{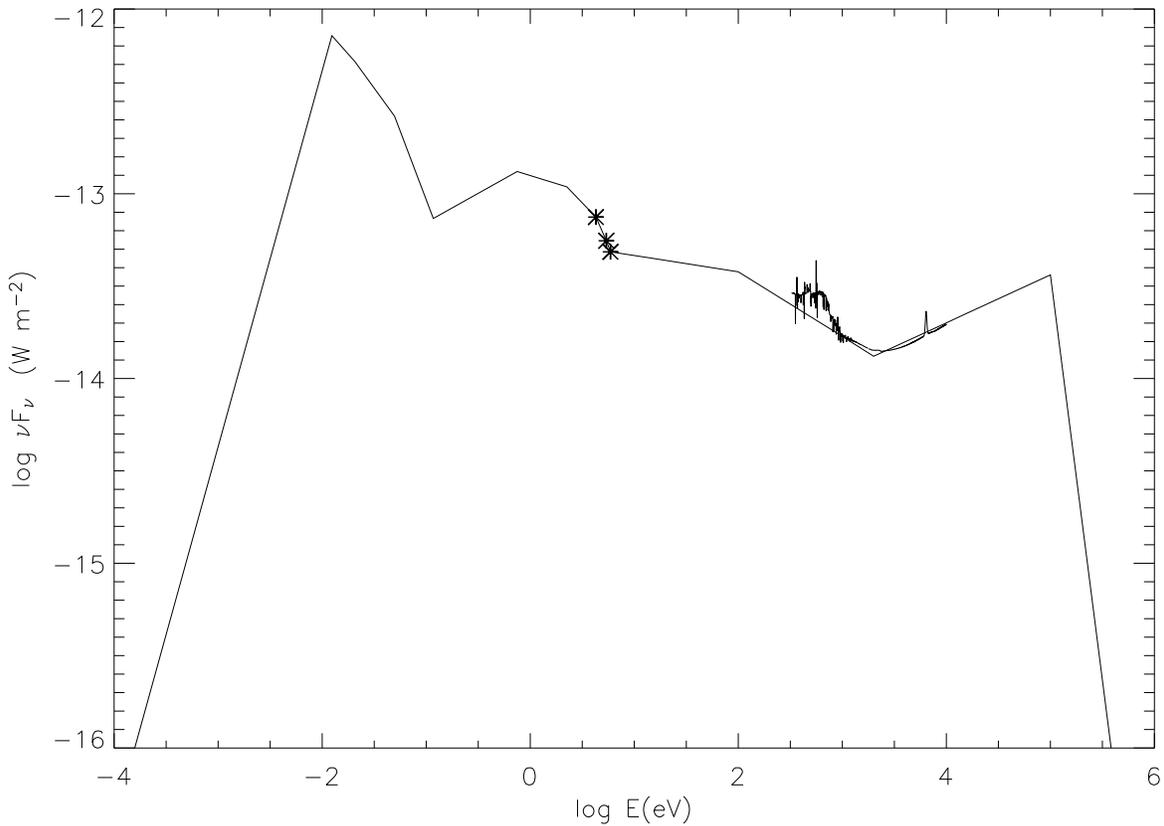}
\figcaption{NGC 4051 high-state spectral energy distribution. Best-fit RGS and EPIC models 
            (corrected for Galactic absorption) are over-plotted. {\it XMM-Newton} OM UV fluxes,
            corrected for Galactic reddening, are indicated with asterisks. Additional radio-optical 
            data points are from NED, and we choose a hard X-ray cut-off at 100 keV. The soft X-ray 
            bump at 0.3-1 keV emits a small fraction of the total AGN luminosity, and is unlikely 
            to be black body emission from the accretion disk. Most of the luminosity from the accretion 
            disk is radiated in the optical-UV big blue bump.}
\end{figure}

\begin{figure}[ht]
\epsscale{0.9}
\plotone{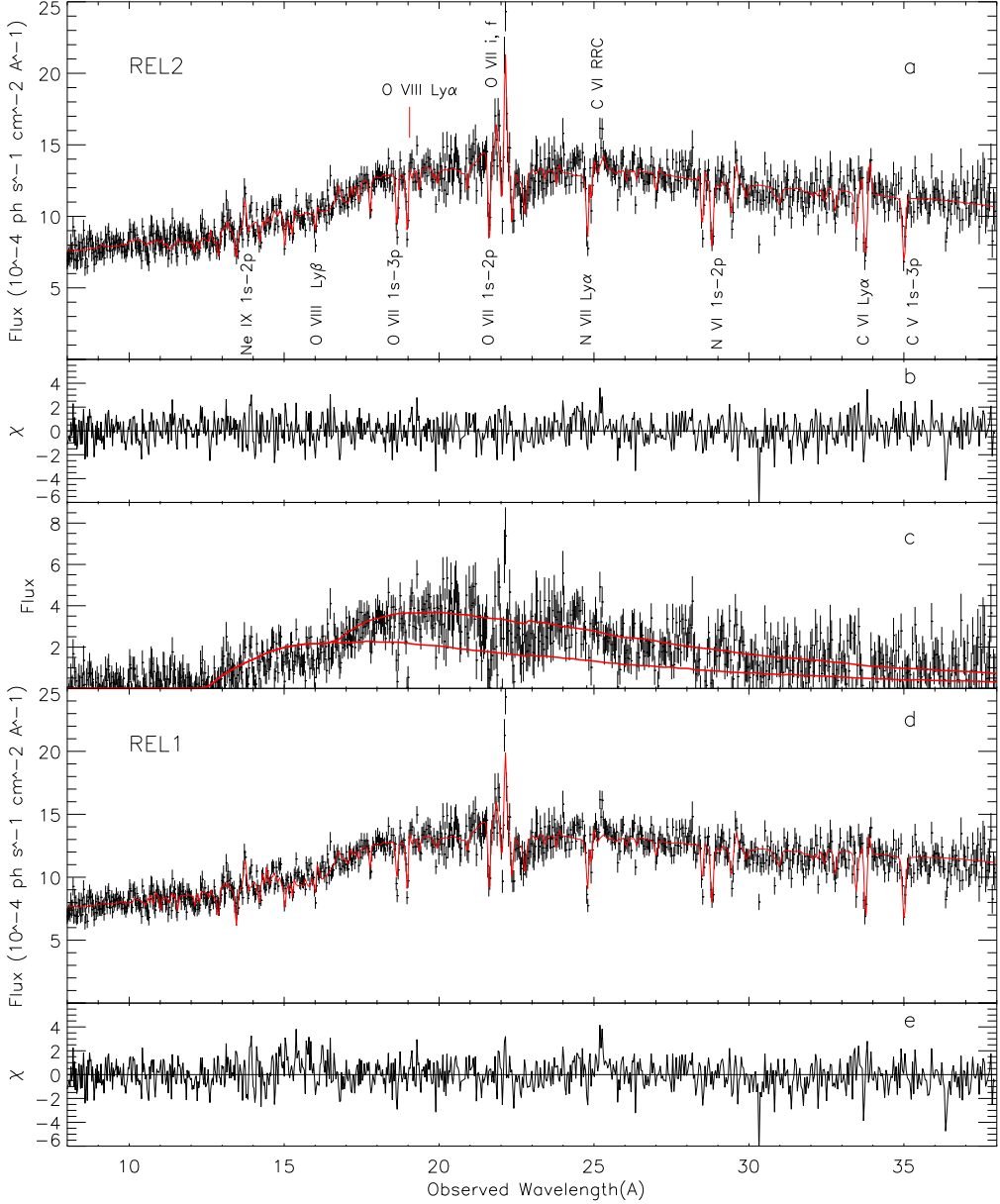}
\figcaption{a) Complete relativistic O {\sc viii} series (REL2) and  d) Ly $\alpha$-only (REL1) 
    model fits to the {\it XMM-Newton} RGS  spectrum of NGC 4051. Both 
    models contain an underlying steep power law continuum ($\Gamma \sim 2.4$). 
    Narrow absorption and emission lines are also included for various ions 
    of C, N, O, Ne, and Fe. REL2 improves upon REL1 by adding in the O {\sc viii} 
    Ly$\beta$ and higher-order relativistic emission lines and RRC. The middle panel 
    (c) shows the residual O {\sc viii} emission after subtracting away the 
     absorbed power-law continuum and narrow emission lines. A large fraction of the 
     relativistic O {\sc viii} flux is from high order lines.}
\end{figure}

\begin{figure}[ht]
\epsscale{1.0}
\plotone{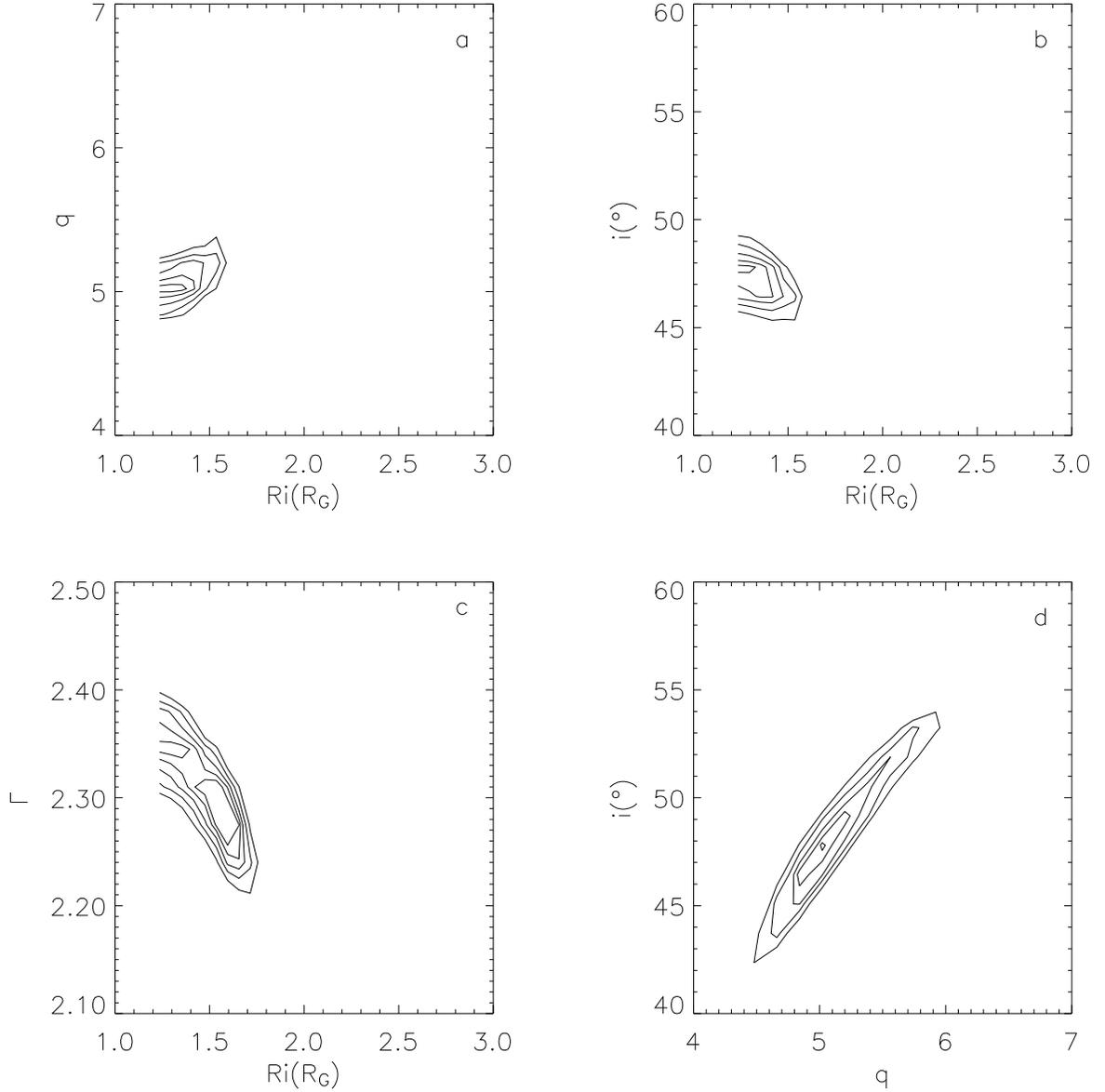}
\figcaption{Confidence contours ($1-5 \sigma$) for relativistic emission line (REL2) fit.
            a-c) The inner radius $R_\mathrm{i}<1.7 R_\mathrm{G}$ is well constrained 
            by the RGS data. a) Radial emissivity index vs. $R_i$. b) Inclination vs. $R_i$.
            c) Power law continuum photon index vs. $R_i$, showing dependence of
               disk inner radius on model continuum shape. 
            d) There is some degeneracy between disk inclination angle $i$ and 
               emissivity index $q$, but they are still well constrained by the
               REL2 model.}
\end{figure}
 
\begin{figure}[ht]
\plotone{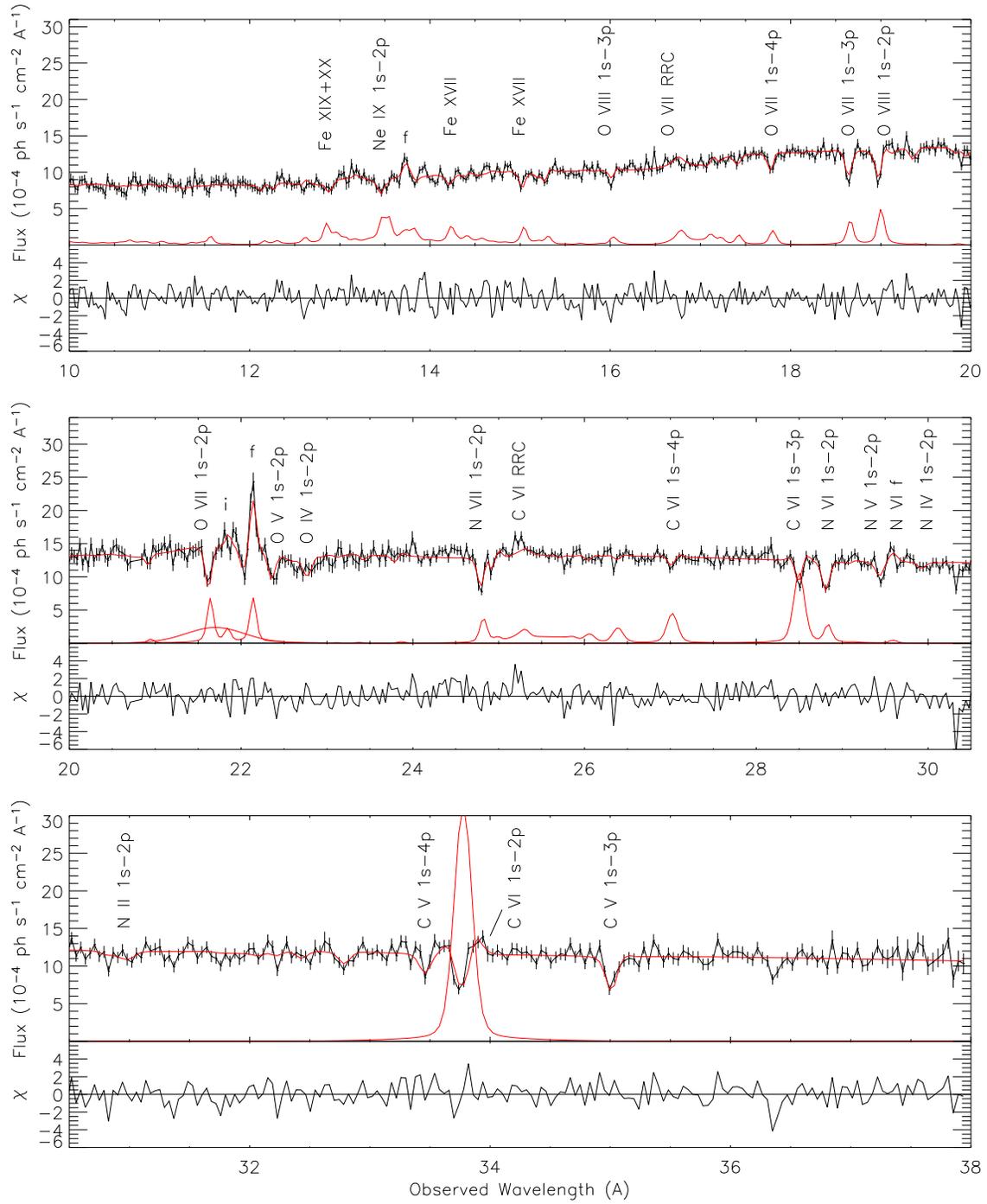}
\epsscale{0.8}
\figcaption{High-state RGS spectrum of ionized absorber-emitter (REL2 model).}
\end{figure}

\begin{figure}[ht]
\epsscale{1.0}
\plotone{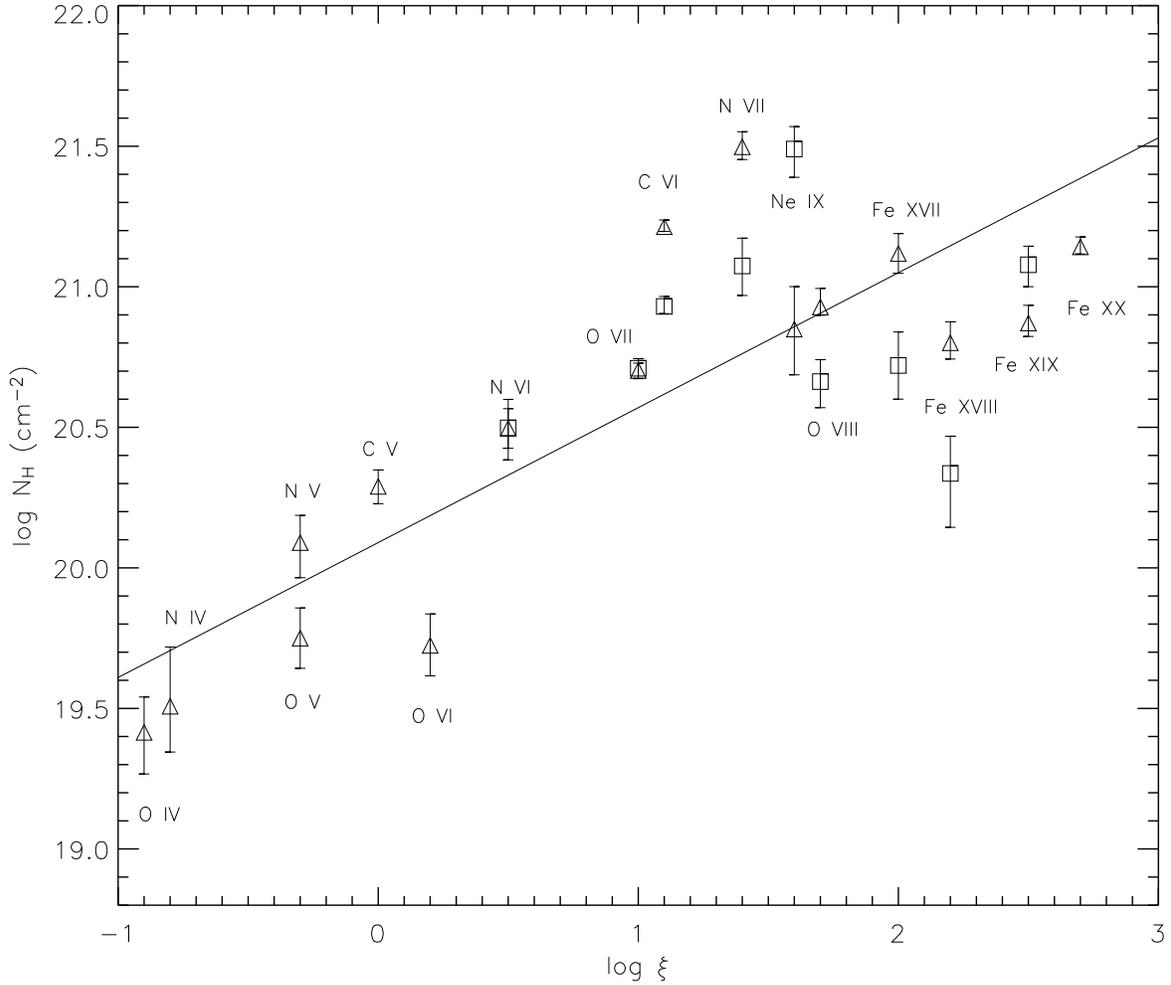}
\figcaption{Ionized absorber-emitter column density distribution. Equivalent hydrogen column 
            densities $N_\mathrm{H}$ for absorber (triangles) and emitter (squares) are
            plotted vs. ionization parameter $\xi$. There is an overall trend of increasing 
            column density with ionization parameter. A power law fit is indicated by the solid 
            line. The observed distribution appears to peak at $\log \xi \sim 1.4$.}
\end{figure}

\begin{figure}[ht]
\plotone{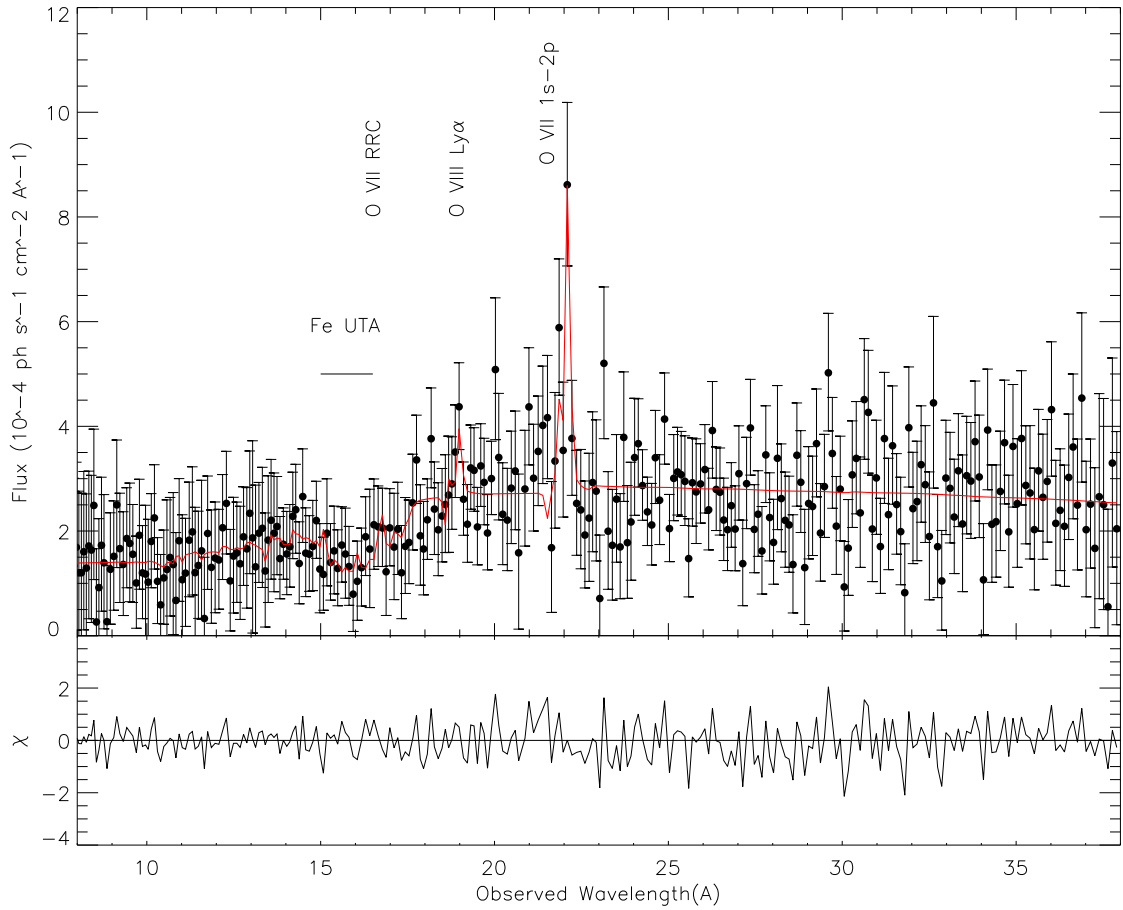}
\figcaption{Model fit to the {\it XMM-Newton} RGS low-state spectrum of NGC 4051. The 
            continuum is well described by a power law with $\Gamma=2.6$. Adding relativistic
            O {\sc viii} emission does not improve the fit. Note the large EW O {\sc vii} narrow 
            line and possible Fe UTA.}
\end{figure}

\begin{figure}[ht]
\plotone{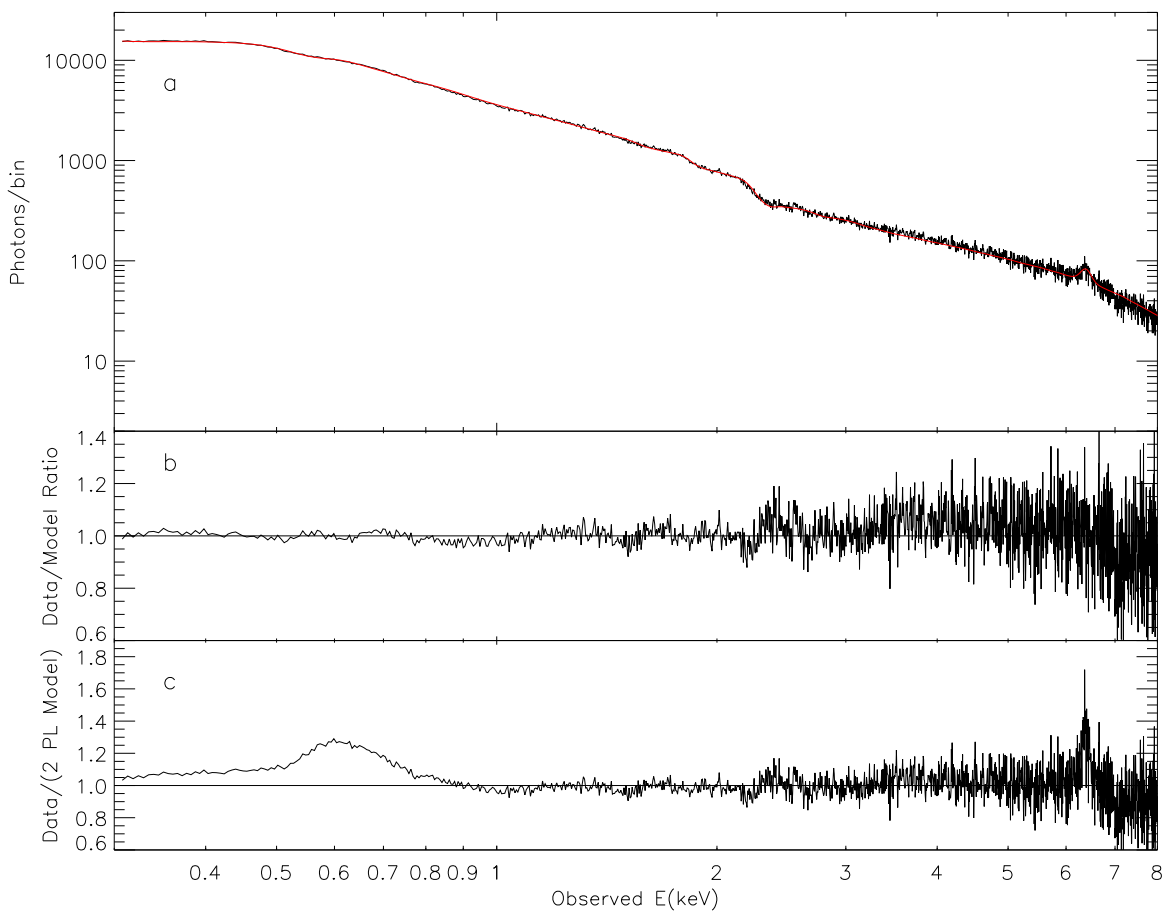}
\figcaption{a) EPIC PN spectrum and model. The model consists of hard and soft power-law continua 
            plus relativistic O {\sc viii} and Fe {\sc xxv} emission lines. Narrow O {\sc vii} and 
            Fe K$\alpha$ emission lines are also included. b) Ratio of data to model. 
            c) Ratio of data to hard plus soft power law continua with Galactic absorption. The large
            bump at 0.6 keV is a combination of relativistic O {\sc viii} and narrow O {\sc vii}
            emission lines. The narrow line at 6.3 keV is low-ionization Fe K$\alpha$.}
\end{figure}

\begin{figure}[ht]
\plotone{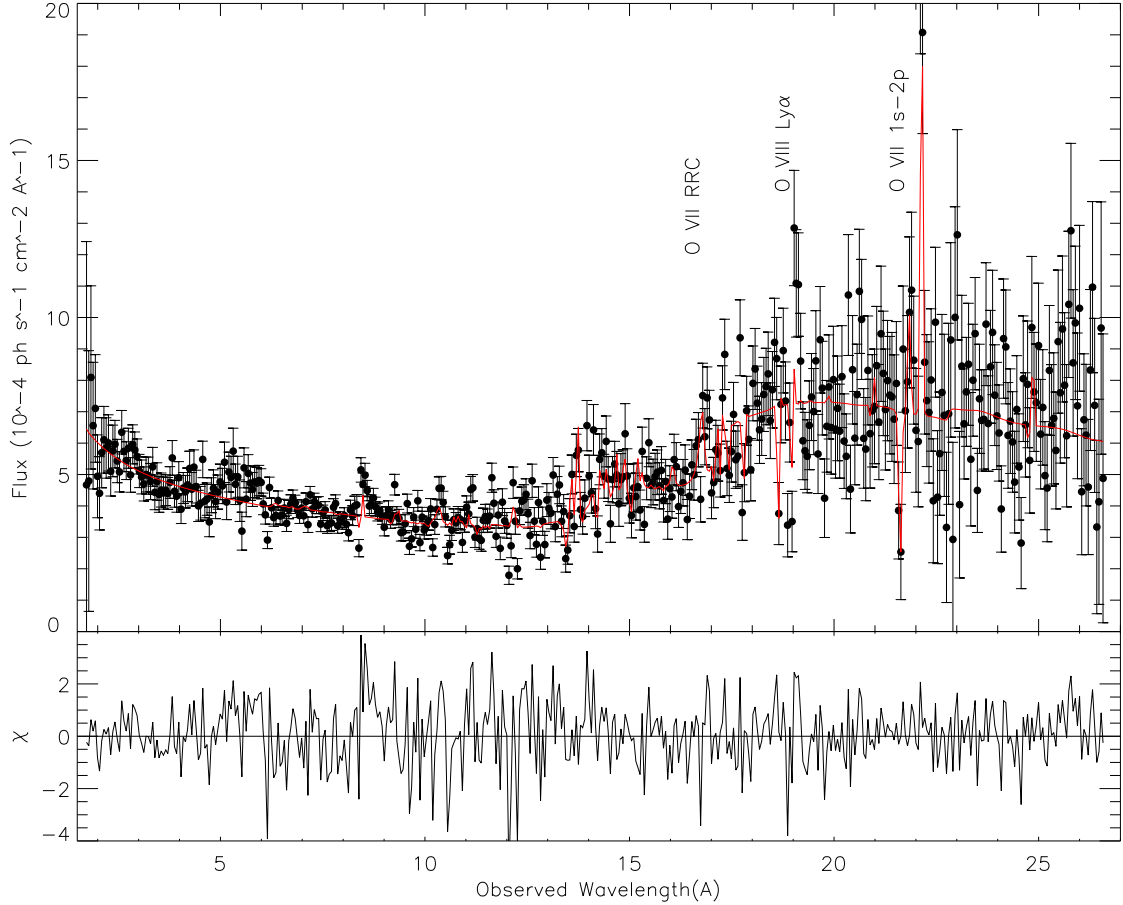}
\figcaption{{\it Chandra} HETG (MEG) spectrum of NGC 4051. The underlying continuum is fit by hard 
            plus soft power laws, and the broad bump in the soft excess is fit by relativistically 
            broadened O {\sc viii} emission. Note the secondary excess at 13-17 \AA, corresponding 
            to blended high-order relativistic emission lines. Emission and absorption from the 
            ionized outflow are also included in the model.}
\end{figure}

\begin{figure}[ht]
\plotone{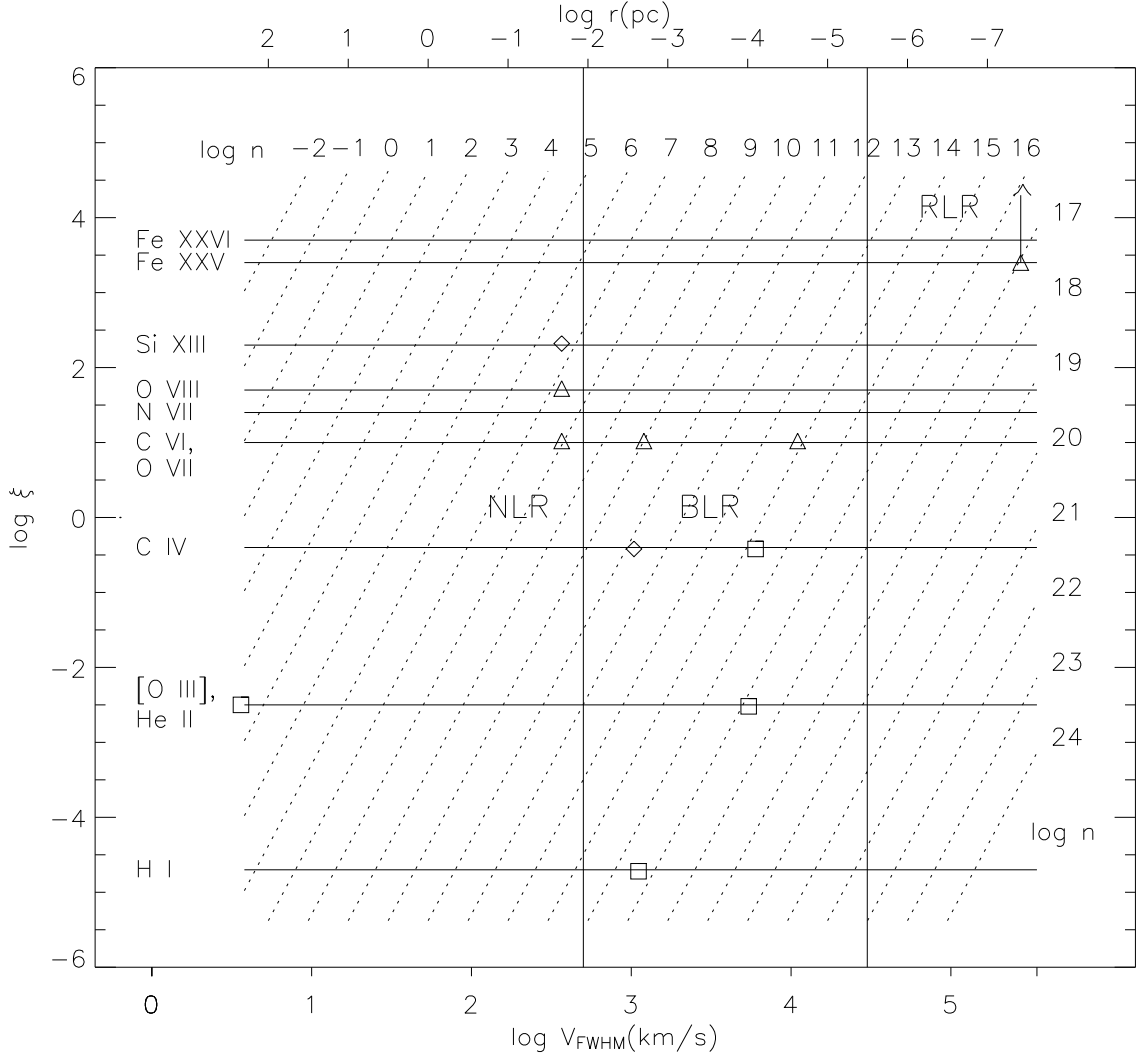}
\figcaption{Ionization map of NGC 4051  NLR, BLR, and RLR. Dotted diagonal lines 
            give ionization parameter $\xi$ vs. Keplerian velocity for densities in the range 
            $\log n (\mathrm{cm}^{-3})=-2$ to 24 and $\log L_\mathrm{ion}(\mathrm{erg~s}^{-1})=42.6$. 
            Velocity, from measured line FWHM, is converted to radial distance assuming isotropic
            circular orbits around a $5 \times 10^5 M_{\odot}$ black hole. The exceptions are narrow 
            [O {\sc iii}] and relativistic O {\sc viii} velocities computed from emitter radii. 
            Vertical lines split the emission line regions by velocity into NLR, BLR, and RLR. 
            Horizontal lines give ionization parameter of peak emissivity for prominent emission 
            lines. Data are plotted as triangles ({\it XMM-Newton}, this work), diamonds 
            \citep{cbk01}, and squares \citep{p00}.}
\end{figure}

\begin{figure}[ht]
\plotone{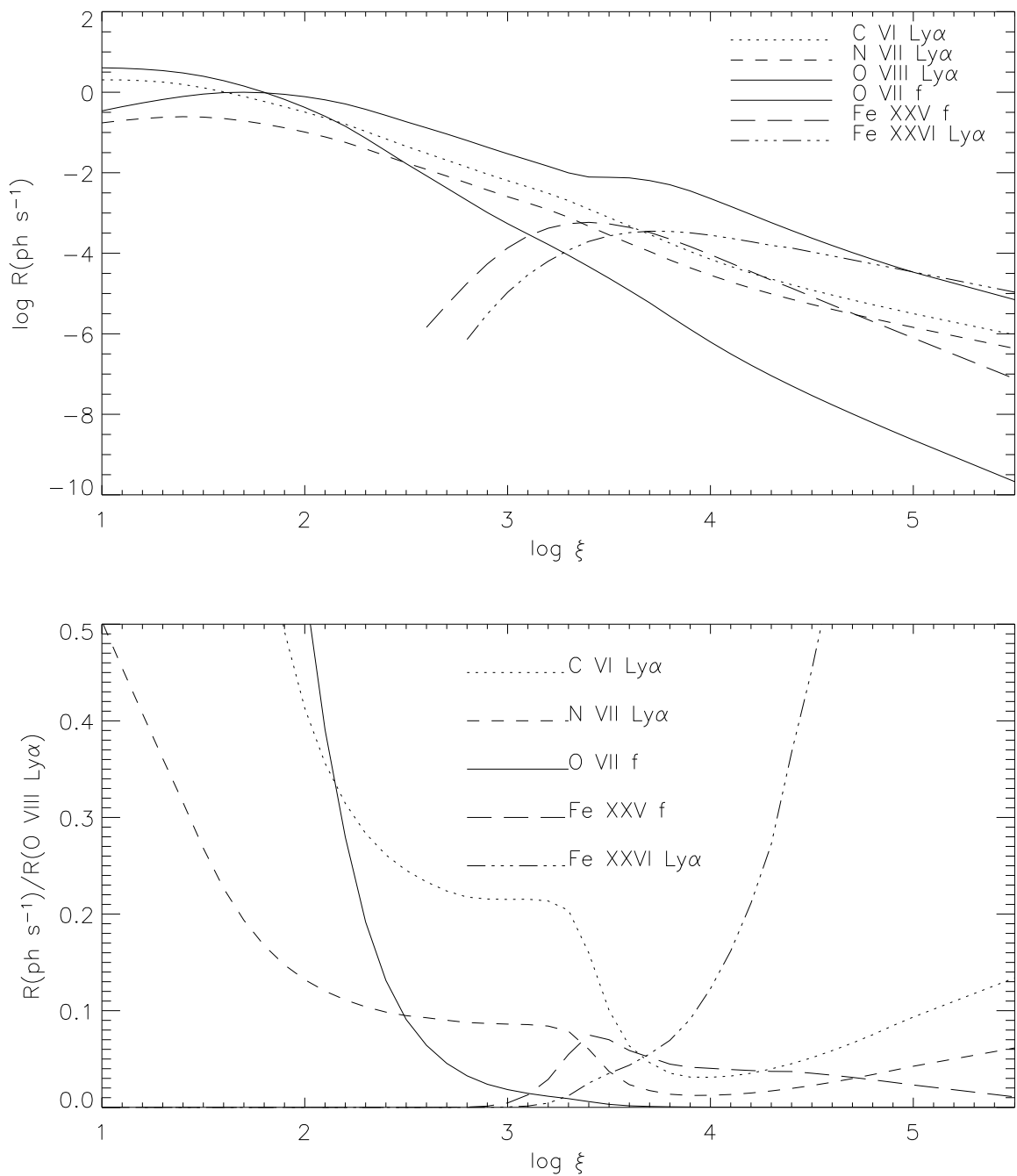}
\figcaption{Emission lines from an optically thin photoionized plasma with solar abundances. 
            a) Selected H-like and He-like $n=2-1$ line strengths are plotted as a 
            function of ionization parameter. b) Ratio of line strengths to O {\sc viii} 
            Ly$\alpha$. The RLR must be located at $\log \xi > 3.4$ to explain the lack
            of C {\sc vi} emission. Fe {\sc xxv} and {\sc xxvi} are the dominant Fe charge
            states at this ionization parameter.
            }
\end{figure}

\begin{figure}[ht]
\plotone{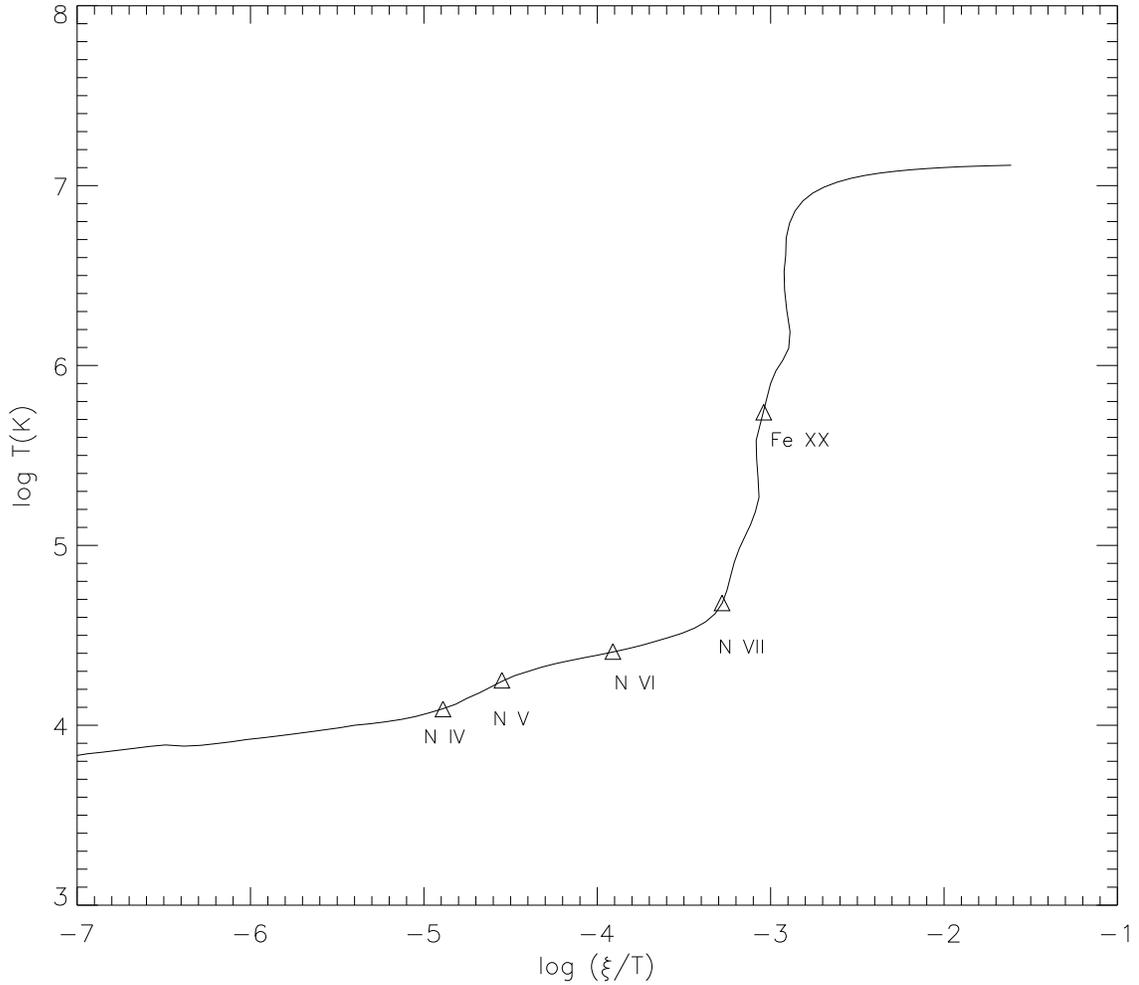}
\figcaption{Thermal equilibrium (heating) curve for photoionized plasma, computed with XSTAR and 
            NGC 4051 SED (Fig. 2). N {\sc iv}-{\sc vii} and Fe {\sc xx} absorbers are indicated,
            assuming peak ionic abundances. The ordinate $\xi/T \sim F_\mathrm{ion}/P$ is
            inversely proportional to the gas pressure, which {\it decreases} to the right. If all 
            ions are located at the same radius, the ionizing flux $F_\mathrm{ion}$ is constant. The 
            observed ionization states cover a factor of 100 in pressure, and are {\it not} all in 
            pressure equilibrium.}
\end{figure}

\begin{deluxetable}{llll}
\tablecaption{High state absorber and emitter column densities}
\tablewidth{0pt}
\tablehead{
\colhead{Ion} & \colhead{$N_\mathrm{abs}$\tablenotemark{a}}  & 
\colhead{$N_\mathrm{em}$\tablenotemark{b}} & 
\colhead{$N_\mathrm{H,abs}$\tablenotemark{c}}}

\startdata
C {\sc v}     & $5^{+1}_{-1}$ & \nodata   & 2 \\
C {\sc vi}    & $33^{+3}_{-2}$ & $22^{+3}_{-2}$ & 16 \\
N {\sc ii}    & $0.6^{+0.4}_{-0.3}$ & \nodata   & 0.7 \\
N {\sc iii}   & $<$0.2 & \nodata & $<$0.3 \\
N {\sc iv}    & $0.2^{+0.2}_{-0.1}$ & \nodata   & 0.3 \\
N {\sc v}     & $0.5^{+0.2}_{-0.2}$ & \nodata   & 1\\
N {\sc vi}    & $2.5^{+0.7}_{-0.6}$ & $1.9^{+0.8}_{-0.7}$ & 3.1 \\
N {\sc vii}   & $19^{+4}_{-3}$ & $9^{+4}_{-3}$ & 31 \\
O {\sc iv}    & $1.3^{+0.7}_{-0.6}$ & \nodata   & 0.26 \\
O {\sc v}     & $2.0^{+0.9}_{-0.7}$ & \nodata   & 0.56 \\
O {\sc vi}    & $1.7^{+0.8}_{-0.6}$ & \nodata   & 0.53 \\ 
O {\sc vii}   & $30^{+3}_{-3}$ & $23^{+3}_{-3}$ & 5.0 \\
O {\sc viii}  & $38^{+10}_{-4}$ & $27^{+9}_{-8}$  & 8.5 \\
Ne {\sc ix}   & $6^{+4}_{-3}$ & $13^{+4}_{-4}$  & 7 \\
Fe {\sc i}    & $<$0.2  & \nodata & $<$2 \\
Fe {\sc xvii} & $2.1^{+0.6}_{-0.5}$ & $0.8^{+0.4}_{-0.3}$  & 13 \\
Fe {\sc xviii}& $1.0^{+0.3}_{-0.2}$ & $0.4^{+0.2}_{-0.2}$  & 6.3 \\ 
Fe {\sc xix}  & $1.2^{+0.3}_{-0.2}$ & $1.9^{+0.5}_{-0.5}$  & 7.4 \\
Fe {\sc xx}  &  $2.2^{+0.3}_{-0.2}$ & $5^{+1}_{-1}$  & 14 \\
\enddata

\tablenotetext{a}{Absorber ionic column density ($10^{16}$ cm$^{-2}$)}
\tablenotetext{b}{Emitter ionic column density ($10^{16}$ cm$^{-2}$)}
\tablenotetext{c}{Equivalent absorber hydrogen column density ($10^{20}$ cm$^{-2}$),
                  assuming solar elemental abundances \citep{angr89}.}

\end{deluxetable}


\begin{thebibliography}{}

\bibitem[Agol \& Krolik (2000)]{ak00} Agol, E., \& Krolik, J. H. 2000, ApJ, 528, 161

\bibitem[Anders \& Grevesse (1989)]{angr89} Anders, E., \& Grevesse, N. 1989, 
Geochimica et Cosmochimica Acta, 53, 197

\bibitem[Arav et al. (2003)]{aks03} Arav, N., Kaastra, J., Steenbrugge, K., 
Brinkman, B., Edelson, R., Korista, K. T., \& de Kool, M. 2003, ApJ, 590, 174

\bibitem[Ballantyne, Iwasawa \& Fabian (2001)]{bif01} Ballantyne, D. R.,
Iwasawa, K. \& Fabian, A. C. 2001, MNRAS, 323, 506 

\bibitem[Ballantyne, Ross, \& Fabian (2002)]{brf02} Ballantyne, D. R., 
  Ross, R. R., \& Fabian A. C., 2002, MNRAS, 336, 867

\bibitem[Ballantyne, Vaughan, \& Fabian (2003)]{bvf03} Ballantyne, D. R., 
  Vaughan, S., \& Fabian A. C., 2003, MNRAS, 342, 239

\bibitem[Behar et al. (2003)]{brb03} Behar, E. et al. 2003, ApJ, 598, 232

\bibitem[Behar \& Netzer (2002)]{bn02} Behar, E. \& Netzer, H. 2002, ApJ, 570, 
165

\bibitem[Boller, Brandt, \& Fink (1996)]{bbf96} Boller, Th., Brandt, W. N., 
\& Fink, H. 1996, A\&A, 305, 53

\bibitem[Boller et al. (2003)]{btf03} Boller, Th., Tanaka, Y., Fabian, A., 
Brandt, W. N., Gallo, L., Anabuki, N., Haba, Y., \& Vaughan, S. 2003, MNRAS, 
L343, 89

\bibitem[Brookings \& Ogle (2003)]{bo03} Brookings, T., \& Ogle, P. M. 2003
{\it IMP--IDL Multi-Purpose Fitting User Guide}

\bibitem[Branduardi-Raymont et al. (2001)]{br01} Branduardi-Raymont, G.,
Sako, M., Kahn, S., Brinkman, A. C., Kaastra, J. S., \& Page, M. J.  2001, 
A\&A, L140

\bibitem[Cash (1979)]{c79} Cash, 1979, ApJ, 228, 939

\bibitem[Christopoulou et al. (1997)]{chs97} Christopoulou, P. E., Holloway, 
A. J., Steffen, W., Mundell, C. G., Thean, A. H. C., Goudis, C. D., Meaburn, J.,
\& Pedlar, A. 1997, MNRAS, 284, 385

\bibitem[Collinge et al. (2001)]{cbk01} Collinge, M. J., et al. 2001, ApJ,
557, 2

\bibitem[den Herder et al. (2001)]{dh01} den Herder, J. W., et al. 2001, A\&A, 365, L7 

\bibitem[Dickey \& Lockman (1990)]{dl90} Dickey, J. M. \& Lockman, F. J. 1990, ARAA, 28, 215
 
\bibitem[Fabian et al. (1989)]{frs89} Fabian, A. C., Rees, M. J., Stella, L., 
\& White, N. E., MNRAS, 238, 729

\bibitem[Fabian et al. (1995)]{fnr95} Fabian, A. C., Nandra, K., Reynolds, C. S., Brandt, 
W. N., Otani, C., \& Tanaka, Y. 1995, MNRAS 277, L11

\bibitem[Gehrels (1986)]{g86} Gehrels, N. 1986, ApJ, 303, 336

\bibitem[Goodrich (1989)]{g89} Goodrich, R. W. 1989, ApJ, 342, 224

\bibitem[Guainazzi et al. (1996)]{g96} Guainazzi, M., Mihara, T., Otani, C., \&
Matsuoka, M. 1996, PASJ, 48, 781

\bibitem[Gu (2002)]{g02} Gu, M. F., 2002, {\it FAC 0.8.8 Manual}

\bibitem[Kaastra et al.(2002)]{ksr02} Kaastra, J. S., Steenbrugge, K. C., Raassen, A. J. J.,
van der Meer, R. L. J., Brinkman, A. C., Liedahl, D. A., Behar, E., \& de Rosa, A. 
2002, A\&A, 386, 427

\bibitem[Kallman \& McCray (1982)]{km82} Kallman, T. R., \& McCray, R., 1982, ApJS, 50, 263

\bibitem[Kallman (2002)]{k02} Kallman, T. R. 2002, {\it XSTAR, A Spectral Analysis Tool, V2.1}

\bibitem[Kirsch et al. (2003)]{cal03} Kirsch, M., et al. 2003, Document XMM-SOC-CAL-TN-0018, ESA

\bibitem[Krolik \& Kriss (2001)]{kk01} Krolik, J. H., \& Kriss, G. A. 2001, ApJ, 561, 684

\bibitem[Krolik (2001)]{k01} Krolik, J. H. 2001, ApJ, 551, 72

\bibitem[Krongold et al. (2003)]{ke03} Krongold, Y., Nicastro, F., Brickhouse, N. S., Elvis, M.,
Liedahl, D. A., \& Mathur, S., 2003, ApJ, 597, 832

\bibitem[Laor (1991)]{l91} Laor, A. 1991, ApJ, 376, 90

\bibitem[Laor (2003)]{l03} Laor, A. 2003, ApJ, L590, 86

\bibitem[Lee et al. (2001)]{loc01} Lee, J.C., Ogle, P.M., Canizares, C. R.,
Marshall, H. L., Schulz, N. S., Morales, R., Fabian, A. C., \& Iwasawa, K.
2001, ApJ, 554, L13

\bibitem[Leighly (2000)]{l00} Leighly, K. M. 2000, NewAR, 44, 395

\bibitem[Leighly (1999a)]{l99a} Leighly, K. M. 1999, ApJS, 125, 297

\bibitem[McHardy et al. (1995)]{mgd95} McHardy, I. M., Green, A. R., Done, D., Puchnarewicz, 
E. M., Mason, K.O., Branduardi-Raymont, G., \& Jones, M. H. 1995, MNRAS, 273, 549

\bibitem[McKernan et al. (2003)]{myg03} McKernan, B., Yaqoob, T., George, I. M., \& Turner,
T. J. 2003, ApJ, 593, 142

\bibitem[Mason et al. (2002)]{mmp02} Mason, K. O., et al. 2002, ApJ, L580, 117

\bibitem[Mason et al. (2003)]{mbo03} Mason, K. O., et al. 2003, ApJ, 582, 95 

\bibitem[Murray \& Chiang (1997)]{mc97} Murray, N., \& Chiang, J. 1997, ApJ, 474, 91

\bibitem[Nandra et al (1997a)]{ngm97a} Nandra, K., George, I. M., Mushotzky, R. F., Turner, 
T. J., \& Yaqoob, T. 1997a, ApJ, 476, 70 

\bibitem[Nandra et al (1997b)]{ngm97b} Nandra, K., George, I. M., Mushotzky, R. F., Turner, 
T. J., \& Yaqoob, T. 1997b, ApJ, 477, 602 

\bibitem[Nayakshin, Kazanas, \& Kallman (2000)]{nkk00} Nayakshin, S., Kazanas, D., \&
Kallman, T. R. 2000, ApJ, 537, 833 

\bibitem[Netzer \& Laor (1993)]{nl93} Netzer, H., \& Laor, A., 1993, ApJ, 404, L51

\bibitem[Novikov \& Thorne (1973)]{nt73} Novikov, I. D., \& Thorne, K. S. 1973. in
{\it Black Holes}, ed. C. DeWitt \& B. DeWitt, (New York: Gordon and Breach).

\bibitem[Nowak \& Chiang (2000)]{nc00} Nowak , M. A., \& Chiang, J. 2000, ApJ, 531, L13

\bibitem[Ogle et al. (2000)]{o00} Ogle, P. M., Marshall, H. L., Lee, J. C.,
\& Canizares, C. R. 2000, ApJ, L81

\bibitem[Ogle et al. (2003)]{obc03} Ogle, P. M., Brookings, T., Canizares, 
  C. R., Lee, J. C., \& Marshall, H. L. 2003, A\&A, 402, 849

\bibitem[Ogle \& Brookings (2003)]{ob03} Ogle, P. M., \& Brookings, T. 2003
{\it IMP--IDL Multi-Purpose Fitting Spectroscopy Guide}

\bibitem[Page et al. (2003)]{p03} Page, M. J., Mason, K. O., Uttley, P., McHardy, I. M., \& Ogle, 
P. M., MNRAS, in prep. 

\bibitem[Page et al (2001)]{pmc01} Page, M. J., et al. 2001, A\&A, 365, 152

\bibitem[Peterson et al. (2000)]{p00} Peterson, B. M. et al. 2000, ApJ, 542, 161

\bibitem[Pounds et al. (2003)]{prp03} Pounds, K. A., Reeves, J. N., Page, K. L., 
Wynn, G. A., \& O'Brien, P. T. 2003, MNRAS, 342, 1147

\bibitem[Porquet \& Dubau (2000)]{pd00} Porquet, D., \& Dubau, J. 2000, A\&AS, 143, 495

\bibitem[Reynolds et al. (1997a)]{rwf97} Reynolds, C. S., Ward, M. J., Fabian, A. C., \& Celotti, A.
1997a, MNRAS, 291, 403 

\bibitem[Reynolds \& Begelman (1997b)]{rb97} Reynolds, C. S., \& Begelman, M. C. 1997b, 
ApJ, 488, 109

\bibitem[Rodriguez-Pascual et al. (1997)]{rp97} Rodriguez-Pascual, P. M., Mas-Hesse, J. M., 
\& Santos-Lleo, M. 1997, A\&A, 327, 72

\bibitem[Sako et al. (2003a)]{skb03} Sako, M., et al. 2003a, ApJ, 596, 114

\bibitem[Sako (2003b)]{s03} Sako, M. 2003b, ApJ, 594, 1108

\bibitem[Salvi et al. (2003a)]{salv03a} Salvi, N. J,  et al. 2003a, MNRAS, in press

\bibitem[Salvi et al. (2003b)]{spw03} Salvi, N. J., Mason, K. O., Ogle, P. M. 
et al. 2003b, MNRAS, in prep.

\bibitem[Salvi (2003)]{salv03} Salvi, N. J. 2003, PhD Thesis, University of London

\bibitem[Sambruna et al. (2001)]{snk01} Sambruna, R. M., Netzer, H.,
Kaspi, S., Brandt, W. N., Chartas, G., Garmire, G. P., Nousek, J. A., \& Weaver, K. A. 2001, 
ApJ, 546, L13

\bibitem[Shakura \& Sunyaev (1973)]{ss73} Shakura, N. I., \& Sunyaev, R. A. 1973, A\&A, 24, 337 

\bibitem[Shemmer et al. (2003)]{sun03} Shemmer, O., Uttley, P., Netzer, H., \& McHardy, I. M.
2003, MNRAS, 343, 1341

\bibitem[Socrates, Davis, \& Blaes (2003)]{sdb03} Socrates, A., Davis, S. W., \& Blaes, O. 
2003, ApJ, in press (astro-ph/0307158)

\bibitem[Steenbrugge et al (2003)]{skv03} Steenbrugge, K. C., Kaastra, J. S., de Vries, C. P., 
\& Edelson, R. 2003, A\&A, 402, 477

\bibitem[Tanaka et al. (1995)]{t95} Tanaka, Y., et al. 1995, Nature, 375, 659

\bibitem[Titarchuk (1994)]{tit94} Titarchuk, L., 1994, ApJ, 434, 570

\bibitem[Turner et al. (2001)]{tgy01} Turner, T. J., et al. 2001, ApJ , 548, L13

\bibitem[Uttley et al. (2003a)]{utm03} Uttley, P. Taylor, R. D., McHardy, I. M., Page, M. J., 
Mason, K. O., Lamer, G., \& Fruscione, A. 2003, MNRAS, in press (astro-ph/0310701)

\bibitem[Uttley et al. (2003b)]{ufh03} Uttley, P., Fruscione, A., McHardy, I., \& Lamer, G. 2003, 
ApJ, 595, 656

\bibitem[Vaughan et al. (1999)]{vpr99} Vaughan, S., Pounds, K. A., Reeves, J. Warwick, R., \& Edelson, R. 1999,
MNRAS, 308, L34

\bibitem[Verheijen \& Sancisi (2001)]{vs01} Verheijen, M. A. W., \& Sancisi, R. 2001, A\&A 370, 765

\bibitem[Verner et al. (1996a)]{v96a} Verner, D. A., Verner, E. M., 
\& Ferland, G. I 1996a, Atomic Data Nucl. Data Tables, 64, 1
    
\bibitem[Verner et al. (1996a)]{v96b} Verner, D. A., Ferland, G. J., 
\& Yakovlev, D. G., 1996b, ApJ, 465, 487

\bibitem[Wilms et al. (2001)]{wrb01} Wilms, J., Reynolds, C. S., Begelman, M. C., Reeves, J.,
Molendi, S., Stuabert, R., \& Kendziorra, E. 2001, MNRAS, 328, L27

\end{thebibliography}
\end{document}